\documentstyle[12pt,epsfig]{article}
\newlength{\aivwidth}   
\newlength{\tmpwidth}   

\newcommand{\lsim}
{\;\raisebox{-.3em}{$\stackrel{\displaystyle <}{\sim}$}\;}
\newcommand{\gsim}
{\;\raisebox{-.3em}{$\stackrel{\displaystyle >}{\sim}$}\;}

\begin{document}

\setcounter{page}{1}
\thispagestyle{empty}

\large
\begin{flushright}
BI-TP 96/41\\
hep-ph/9608453\\
August 1996
\end{flushright}
\vspace*{5mm}
\large
\begin{center}
{\bf Improved Effective Vector-Boson Approximation\\
for Hadron-Hadron Collisions\footnote{
Partially supported by the EC-network contract CHRX-CT94-0576 and the
Bundesministerium f\"{u}r Bildung und Forschung, Bonn, Germany}\\}
\vspace*{10mm}
\large
{I. Kuss\\
{\em Fakult\"at f\"ur Physik, Universit\"at Bielefeld}\\
{\em Postfach 10 01 31, 33501 Bielefeld, Germany}}
\end{center}
\normalsize
\vspace*{8mm}

\begin{abstract}
An improved effective vector-boson approximation is applied
to hadron-hadron collisions. 
I include an approximative treatment and
compare with a complete perturbative
calculation for the specific example of $ZZ$ production. 
The results are also compared with existing approaches.
The effective vector-boson approximation in this form
is accurate enough to reproduce
the complete calculation within 10\%. 
This is true even far away from a possible
Higgs boson resonance
where the transverse intermediate vector-bosons
give the dominant contribution.
\end{abstract}

\section{Introduction}

The effective photon approximation
(EPA) (Weiz\-s\"{a}\discretionary{k-}{k}{ck}er-Williams-approximation)
of QED \cite{epa} has proved to be a useful tool
in the study of photon-photon processes at $e^+e^-$ colliders.
With the prospect of high-energy hadron-hadron colliders,
the possibility to study the scattering of {\em massive} vector-bosons
is given. Massive vector-boson scattering is of particular interest as
the symmetry breaking sector of the electroweak theory and the
self-interactions of the vector-bosons are directly
tested.
The method equivalent to the EPA applying to 
the scattering of massive vector-bosons
is the effective vector-boson approximation (EVBA) \cite{chaga,
dawson,karero,lindfors}. The EVBA can be applied to
fermion-fermion scattering processes
in which the final state consists of two fermions and a state $\Xi$ which
can be produced by the scattering of two vector-bosons.
The fermion-fermion cross-section is written as a product of a probability
distribution and
the cross-sections for vector-boson scattering.
The probability distribution describes
the emission of vector-bosons from fermions. The method is an
approximate one which neglects Feynman diagrams of bremsstrahlung-type.
In general, the method is applicable if the fermion scattering energy 
is large against the masses of the electroweak vector-bosons.

\begin{figure}
\begin{center}
\epsfig{file=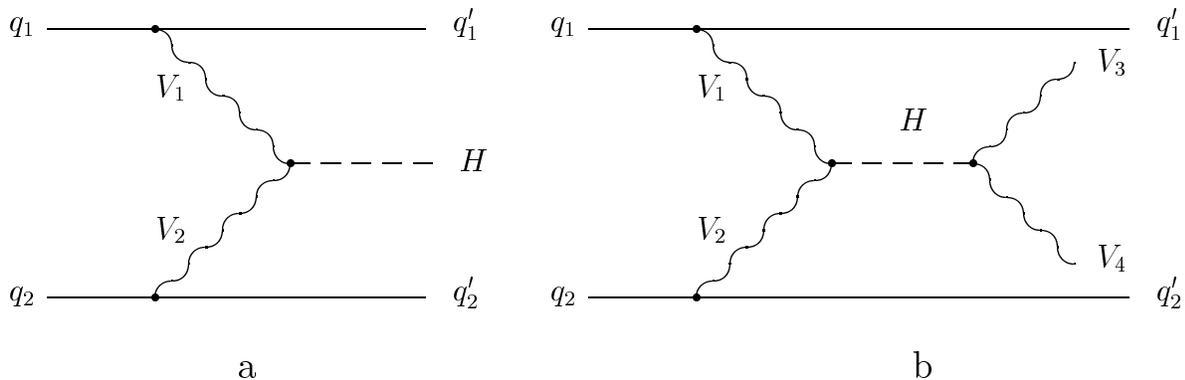,width=16cm,height=5.1cm}
\end{center}
\caption{a: The diagram for Higgs boson production in quark-quark
scattering, $q_1q_2\to q'_1q'_2H$.
b: The diagram for the production of a vector-boson pair $V_3V_4$
as the decay products of a heavy Higgs boson in $q_1q_2$ scattering.}
\label{fig1}
\end{figure}

The possibility of an EVBA has been first noticed in connection with
heavy Higgs boson production \cite{cd}.
The Higgs boson can be produced via the diagram in Figure
\ref{fig1}a, where a sum over all vector-boson pairs $V_1,V_2$ which can
couple to the Higgs boson is to be taken.
In this early application of the EVBA only the contribution from 
longitudinally polarized intermediate vector-bosons, $V_{1,L}V_{2,L}$,
was calculated and the result was found to give a reasonable approximation
to an exact perturbative calculation \cite{cahn}.

Subsequently, the EVBA was also applied to processes of the type
$pp\to V_3V_4+X$,  where two vector-bosons are produced.
The vector-bosons
$V_3$ and $V_4$ emerge as the decay products of a near-resonant heavy
Higgs particle \cite{chaga}.
The scattering process was described by the diagram in Figure \ref{fig1}b.
Also in this case, the inclusion of
only longitudinal intermediate vector-bosons was sufficient. 
It was noted that the production of heavy particles (Higgs bosons or
fermions) is mainly due to the longitudinal intermediate states \cite{karero}.

The concept of vector-bosons as partons in quarks 
was further
established and expressions for vector-boson distributions
in quarks were derived \cite{dawson,karero}. The expressions were given for
all polarizations of the intermediate vector-bosons.
By convolution with the quark distributions in a proton, numerical
results were given for vector-boson distributions in a proton \cite{dawson}.
For the production via two intermediate vector-bosons
it was assumed that convolutions of the distributions
of single vector-bosons could describe the emission probability
of the vector-boson pair.
The EVBA in this form gave reliable results for heavy Higgs boson production
\cite{dawson,karero,cahn,hevyhiggs,dkr} 
and heavy fermion production \cite{hevyfer}.

\begin{figure}
\begin{center}
\epsfig{file=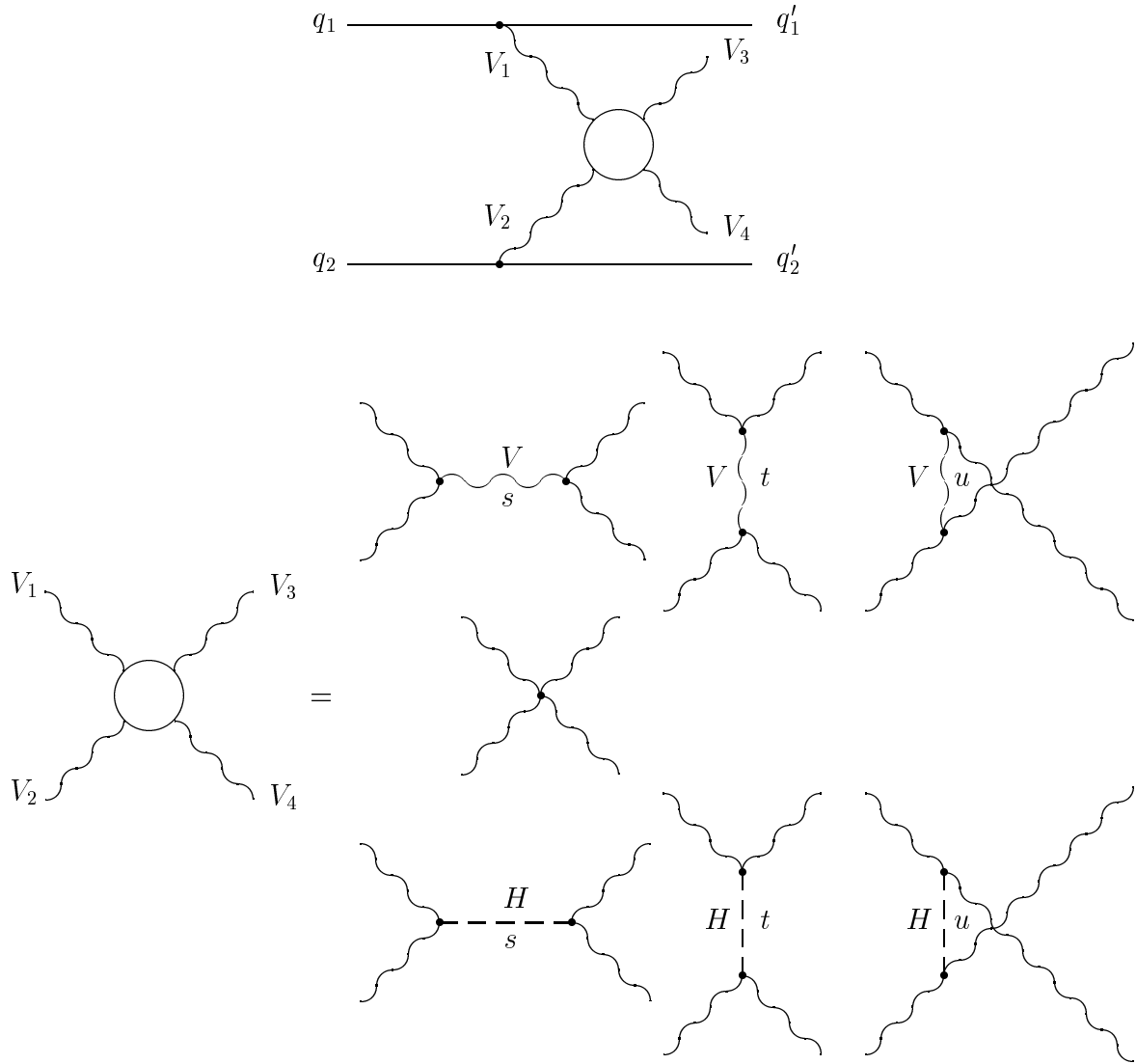,width=16cm,height=14.8cm}
\end{center}
\caption{The diagram for $q_1q_2\to q'_1q'_2 V_3V_4$ in the effective 
vector-boson approximation and the diagrams for vector-boson 
scattering.}
\label{fig2}
\end{figure}

The necessity to include all vector-boson scattering diagrams 
for $V_1 V_2\to V_3 V_4$ in order to obtain EVBA 
predictions for the production of a vector-boson pair $V_3V_4$,
not necessarily near a Higgs boson resonance, has been
first mentioned in \cite{dkr} and \cite{chaga2}. 
The possible diagrams for these processes,
$q_1q_2\to q'_1q'_2 V_3V_4$, where $q_i,q'_i$ are quarks,
are shown in Figure \ref{fig2}.
It was further pointed out that the yield of $V_3V_4$ pairs 
from $q_1q_2\to q'_1q'_2 V_3V_4$
must be discussed together with the yield from the direct reaction
$q_1q_2\to V_3V_4$ (Drell-Yan reaction)
unless a suitable analysis of the different proton
remnants from the two production mechanisms allows to separate
the different production mechanisms. 

In first applications to vector-boson scattering, 
again only the contribution from the longitudinal 
intermediate states
was considered while the contribution from transverse
states was neglected. This contribution was taken to be small
against the $q_1q_2\to V_3V_4$ contribution while the contribution
from $V_{1,L}V_{2,L}\to V_3V_4$ could be large if the longitudinal
vector-bosons interact strongly. 
The interest in the strongly interacting scenario \cite{strong}
was the original motivation to use the EVBA.

\begin{figure}
\begin{center}
\epsfig{file=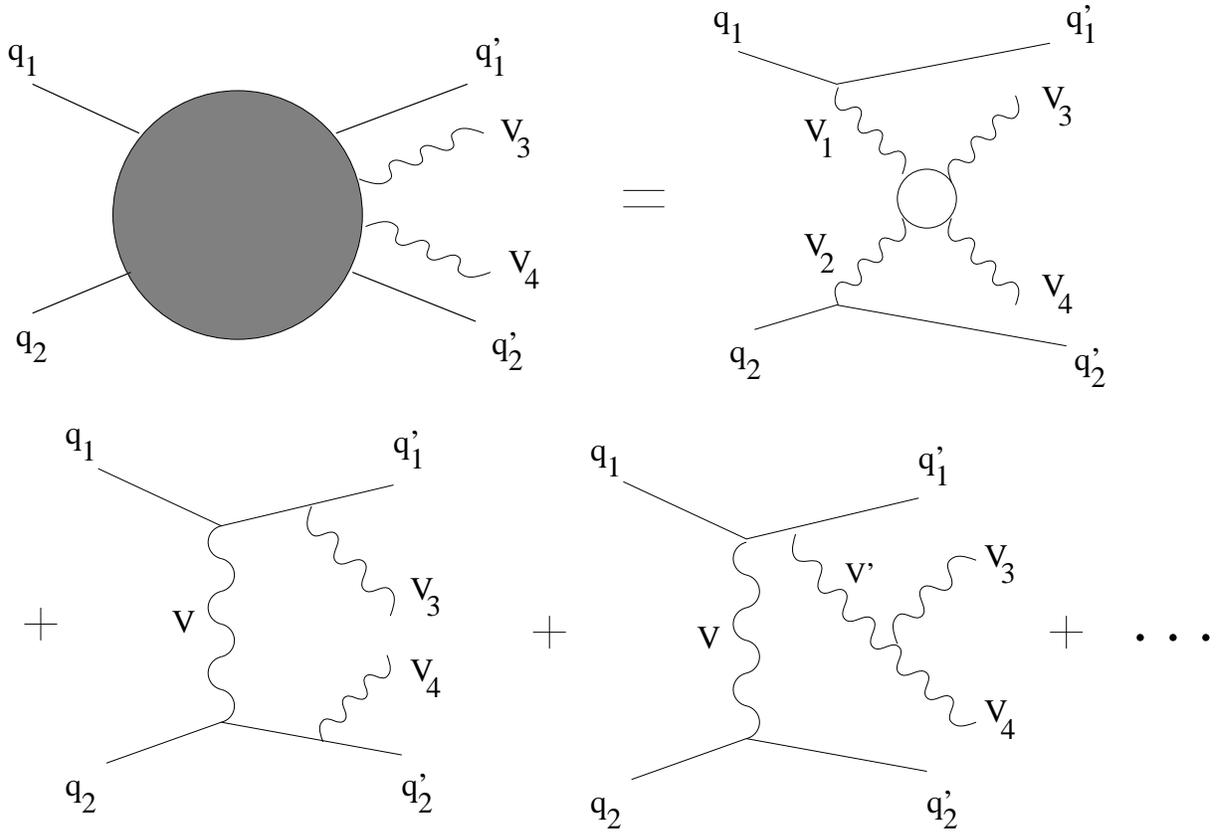,width=16cm,height=11.3cm,angle=0}
\end{center}
\caption{Some of the Feynman diagrams for a process
$q_1 q_2\to q'_1 q'_2 V_3V_4$ in a complete perturbative calculation.
In the top row to the right 
is the diagram for vector-boson scattering, which is
the only type of diagram which is considered in the effective
vector-boson approximation. The bottom row
shows diagrams of bremsstrahlung-type.}
\label{fig3}
\end{figure}

The EVBA has been used for vector-boson scattering in
\cite{chaga}, \cite{dobado}-\cite{amps}.
In \cite{goldplate}, 
the EVBA has been used only for the
longitudinal intermediate states. 
The transverse states have been taken
into account by a complete perturbative calculation 
(to lowest order in the coupling)
of the process
$q_1q_2\to q'_1q'_2 V_3V_4$. This calculation requires the evaluation of
more diagrams than only the vector-boson scattering
diagrams, as indicated in Figure \ref{fig3}. 
To be precise, in \cite{goldplate}
the EVBA has been used only to calculate the difference between 
the cross-sections in a
strongly interacting model and in the standard model with a light Higgs boson.
This difference shows an interesting behavior
in a strongly-interacting scenario and was therefore
considered as a potential signal for strongly interacting vector-bosons.
The difference receives a contribution virtually
only from the longitudinal states.
It was found \cite{goldplate} that this calculation  
agrees with a complete perturbative calculation
to about 10\% (evidenced 
for $W^{\pm}Z$ and $W^{\pm}W^{\pm}$ production) if the
standard model with a heavy Higgs boson is taken as the strongly-interacting
model. I note that for strong scattering a method has been recently
described which does not make use of the EVBA \cite{chano}. 

In \cite{dobado,renard,amps} the application of the EVBA was extended to
the contributions from all intermediate polarization states. It was known,
however, that the EVBA can overestimate results of complete perturbative
calculations by a factor of 3 if the transverse helicities are important
\cite{godol,godbole}. Other comparisons of results of complete calculations
for $pp\to V_3V_4+X$ 
with EVBA results \cite{gunion,ardv}
showed that the EVBA is always a good approximation
on the Higgs boson resonance but in general overestimates the transverse
contribution which is the important one for vector-boson scattering
away from a Higgs boson resonance. The EVBA was found to be
unreliable away from the Higgs boson resonance.
Furthermore,
the EVBA result depends strongly on the details
of the approximations made in deriving the EVBA \cite{godol,godbole}.
In particular, the frequently used leading logarithmic approximation
can overestimate the transverse luminosity by an order of magnitude
\cite{capdequi,reptung}.

We will obtain exact luminosities $\cal L$
for vector-boson pairs in a hadron pair
from an improved formulation of the EVBA, previously introduced for 
fermion-fermion scattering in \cite{lumis}. This formulation
makes no approximation in the integration over the phase space
of the two intermediate vector bosons. The only remaining 
assumption, necessary in an EVBA, 
concerns the off-shell behavior of vector-boson cross-sections.
The formulation, however,
involves multiple numerical integrals
and is thus not very practical in itself. 
However, the exact luminosities $\cal L$ 
form a unique basis to derive well-defined approximations which turn
out to be good. They also serve as a testing ground 
against which existing formulations of the EVBA can be examined.

In Section \ref{sec1} the improved EVBA is applied to hadron collisions 
and numerical results for the exact luminosities
are given. We derive useful approximations to the improved
EVBA.
A comparison with previous formulations of the EVBA is given.
Section \ref{compsec} contains a comparison of 
EVBA results with a complete perturbative calculation for $pp\to ZZ+X$.
Details of various existing formulations of the EVBA are discussed in
Appendix \ref{appa}.

\section{Luminosities for Vector-Boson Pairs in Hadron Pairs\label{sec1}}
\subsection{Improved Effective Vector-Boson Approximation}
\begin{figure}[t]
\begin{center}
\epsfig{file=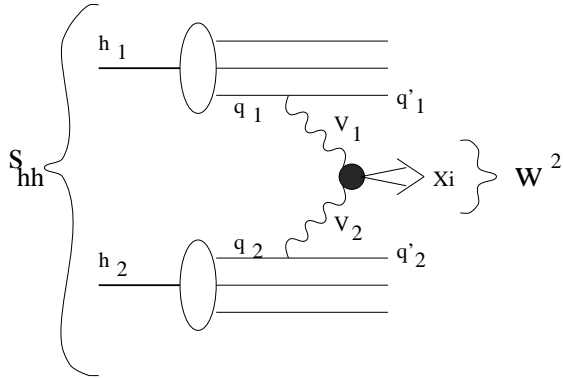,width=7.4cm,height=5cm}
\end{center}
\caption{The diagram for the hadron-hadron scattering process proceeding via
two intermediate vector-bosons,
$h_1h_2\to q_1q_2\to q'_1q'_2V_1V_2$ and $V_1V_2\to \Xi$.}
\label{hhdia}
\end{figure}

Applying the treatment of the improved EVBA \cite{lumis},
we present exact luminosities for finding a vector-boson pair 
in a hadron-pair. 
The luminosities apply to the process
shown in Figure \ref{hhdia}.

The cross-section for
a scattering-process of two hadrons, $h_1$ and $h_2$, with high energies, 
in which an arbitrary final
state $\Xi$ is produced, is given (in the quark-parton model)
by a two-dimensional integral over a product of
parton distribution-functions and the cross-section
for parton-parton scattering processes,
\begin{equation}
\sigma(h_1 h_2\to \Xi +X, s_{hh})=\sum_{q_1,q_2}\int\limits_0^1 
d\xi_1 \int\limits_0^1 d\xi_2
f_{q_1}^{h_1}(\xi_1,\mu_1^2) f_{q_2}^{h_2}(\xi_2,\mu_2^2) 
\sigma(q_1 q_2\to \Xi+X', s_{qq}).
\label{QPM}
\end{equation}
The sum in (\ref{QPM}) extends over all partons (quarks, antiquarks
and gluons) $q_1$ in the hadron $h_1$ and $q_2$ in the hadron $h_2$.
The variable $\xi_i$ in (\ref{QPM}) is the ratio of the momentum of the
parton $q_i$ and the hadron $h_i$.
The quantities
$f_{q_i}^{h_i}(\xi_i,\mu_i^2)$ are the parton distribution functions, evaluated
at the momentum fractions $\xi_i$ and the factorization scales $\mu_i^2$.
The scale is a characteristic energy
of the process which is initiated by the parton $q_i$. 
The quantities $\sigma(q_1 q_2\to \Xi+X')$ in (\ref{QPM})
are the cross-sections for the parton-parton processes.
In (\ref{QPM}), $s_{hh}$ is the square of the hadron-hadron scatterig energy, 
related to the parton-parton scattering energy, $\sqrt{s_{qq}}$,
by 
\begin{equation}
s_{qq}=\xi_1\xi_2s_{hh}.
\end{equation}
The symbols $X$ and $X'$ represent additional particles
in the final state.
In writing down (\ref{QPM}), we assumed that the partons 
have no transverse momentum.
We also neglected the masses of the hadrons and partons. 
These approximations may be made in any frame in which both hadrons are
highly relativistic.

Expressed in terms of the ratio of the squared invariant masses,
$\tau\equiv s_{qq}/s_{hh}$,
and the rapidity $y_q$ of the motion of the center-of-mass of the parton-pair
in the center-of-mass system (cms)
of the hadrons\footnote{The rapidity is taken along the direction of
motion of the hadron $h_1$.}
the cross-section (\ref{QPM}) takes on the form
\begin{eqnarray}
\sigma(h_1h_2\to \Xi+X, s_{hh})&=&\sum_{q_1,q_2}
\int\limits_0^1 d\tau \int\limits_{\frac{1}{2}\ln(\tau)
}^{-\frac{1}{2}\ln(\tau)} 
dy_q\, f_{q_1}^{h_1}\left(\sqrt{\tau}e^{y_q},\mu_1^2\right)
f_{q_2}^{h_2}\left(\sqrt{\tau}e^{-y_q},\mu_2^2\right)\cr
&&\cdot\sigma(q_1 q_2\to \Xi+X', s_{qq}=\tau s_{hh}).
\label{QPM2}\end{eqnarray}
The relations between the variables $\xi_1,\xi_2$ in (\ref{QPM})
and $\tau,y_q$ in (\ref{QPM2}) are given by
$\tau\equiv \xi_1 \xi_2$, $y_q\equiv 1/2\ln\left(
\xi_1/\xi_2\right)$ or, equivalently,  $\xi_1\equiv\sqrt{\tau}e^{y_q}$
and $\xi_2\equiv\sqrt{\tau}e^{-y_q}$.

If the final state $\Xi$ is produced via the vector-boson fusion mechanism,
$V_1V_2\to \Xi$, and the partons are quarks or antiquarks (we will simply
call them quarks here)
an expression for the parton-parton cross-section is given in the EVBA by
\begin{eqnarray}
&&\sigma(q_1 q_2\to \Xi+X',s_{qq})=
\sigma(q_1 q_2\to q'_1 q'_2\Xi, s_{qq})\nonumber\\
&=&\sum_{V_1,V_2}\sum_{pol}\int\limits_0^1 d\hat{x}\,
{\cal L}_{V_1,V_2,pol}^{q_1 q_2}(\hat{x})
\sigma_{pol}(V_1 V_2\to\Xi, {\cal W}^2=\hat{x}s_{qq}).
\label{zwei}\end{eqnarray} 
In (\ref{zwei}), the quantities ${\cal L}_{V_1,V_2,pol}^{q_1 q_2}(\hat{x})$
are luminosities for vector-boson pairs in fermion-pairs. 
The variable $\hat{x}$ is the ratio of the squared invariant mass
${\cal W}^2$ of the vector-boson pair and the one of the quark-pair,
\begin{equation}
\hat{x}\equiv\frac{{\cal W}^2}{s_{qq}}.
\end{equation}
The sum in (\ref{zwei}) runs over all vector-bosons $V_1,V_2$
which can produce
the final state $\Xi$ and over all their helicity states $pol$.
Expressions for the luminosities 
${\cal L}_{V_1,V_2,pol}^{q_1 q_2}(\hat{x})$, 
using no other approximations than those inherent in the effective
vector-boson method, have been given in \cite{lumis}.
The luminosities can be written in the form
\begin{equation}
{\cal L}_{V_1,V_2,pol}^{q_1 q_2}(\hat{x})
=\left(\frac{\alpha}{2\pi}\right)^2
\hat{x}\,c_{q_1(V_1)}^{pol}c_{q_2(V_2)}^{pol}
\int\limits_{\hat{x}}^{1}\frac{d\hat{z}}{\hat{z}}
{\cal L}_{pol}\left(\hat{x},\hat{z},\frac{M_1^2}{s_{qq}},
\frac{M_2^2}{s_{qq}}\right).
\label{wichtig}\end{equation}
In (\ref{wichtig}), $\hat{z}$
is the ratio of the squared invariant mass $M_Y^2$ of a system
consisting of $V_1$ and the quark $q_2$ and the squared invariant
mass of the quark-pair,
\begin{equation}
\hat{z}\equiv\frac{M_Y^2}{s_{qq}}.
\label{zhatdef}\end{equation}
The parameter
$\alpha$ is the fine-structure
constant and $M_i$ are the vector-boson masses.
The quantities
$c_{q_i(V_i)}^{pol}$ are combinations of the vector
and axial-vector couplings, $v_i$ and $a_i$, respectively, of 
$V_i$ to $q_i$. They can be either $v_i^2+a_i^2$
or $2v_ia_i$, depending on $pol$.
In (\ref{wichtig}), the quantities
\begin{eqnarray}
{\cal L}_{pol}\left(\hat{x},\hat{z},\frac{M_1^2}{s_{qq}},
\frac{M_2^2}{s_{qq}}\right)
&=&\eta_0\!\!\!\!\!\int\limits_{-s_{qq}(1-\hat{z})}^0\!\!\!\!\!\!\!dk_1^2
\int\limits_{-s_{qq}\hat{z}(1-\hat{x}/\hat{z})}^0\!\!\!\!\!\!\!\!dK^2_2
\frac{1}{(k_1^2-M_1^2)^2}\frac{1}{(k_2^2-M_2^2)^2}
\int\limits_0^{2\pi}\frac{d\varphi_1}{2\pi} \,
   f_{pol}\,K_{pol}\hspace{2em}
\label{dlumis_amp}
\end{eqnarray}
are ``amputated'' differential luminosities, which do not
anymore contain the fermionic coupling constants.
They depend only on the variables $\hat{x}$ and $\hat{z}$, and, since they are
dimensionless, on the masses of the vector-bosons via the ratios
$M_1^2/s_{qq}$ und $M_2^2/s_{qq}$. In (\ref{dlumis_amp}), $\eta_0$
is the ratio of the on-shell flux factor for the process $V_1V_2\to\Xi$
and the flux factor for the same process evaluated for $M_i^2=0$,
\begin{equation}
\eta_0=\sqrt{ 1 +\left(\frac{M_1^2}{{\cal W}^2}\right)^2
+\left(\frac{M_2^2}{{\cal W}^2}\right)^2
-2\frac{M_1^2}{{\cal W}^2}-2\frac{M_2^2}{{\cal W}^2}
-2\frac{M_1^2}{{\cal W}^2}\frac{M_2^2}{{\cal W}^2} }.
\end{equation}
We simply refer to $\eta_0$ as the on-shell flux factor.
The $k_i^2$ are the squared four-momenta of 
the vector-bosons and 
\begin{equation}
K_2^2\equiv\frac{1}{1-\frac{\textstyle k_1^2}{\textstyle \hat{z}s_{qq}}}k_2^2.
\end{equation}
The quantities $\varphi_1,f_{pol}$ and $K_{pol}$ have been defined in
\cite{lumis}. 
We note that if the momentum of $V_1$ is light-like, $k_1^2=0$, the directions
of motion of $V_1$ and $q_1$ are parallel. In this case,
$\hat{z}$ is the ratio of the energy of $V_1$ and the energy
of $q_1$. The variable $\hat{z}$ has in this case
the same interpretation for the
emission of a vector-boson $V_1$ from a quark $q_1$ as
$\xi_i$ has for the emission of a quark $q_i$ from a hadron $h_i$. 
The corresponding variable for the emission of $V_2$ from $q_2$ is
$\hat{x}/\hat{z}$. 
The vector-bosons
can be approximately treated as ``partons'' in the quarks.
In analogy to $\xi_1$ and $\xi_2$ we introduce
the two variables $\hat{z}_1=\hat{z}$ and $\hat{z}_2=\hat{x}/\hat{z}$.

Inserting (\ref{zwei}) into (\ref{QPM2}) yields
an expression for the cross-section for the production of the state $\Xi$ 
in the hadron-hadron process, proceeding via vector-boson fusion,
\begin{eqnarray}
&&\sigma(h_1 h_2\to q_1 q_2\to q'_1 q'_2 \Xi,s_{hh})\nonumber\\
&\equiv&\sigma(h_1 h_2\to V_1 V_2\to \Xi,s_{hh})\nonumber\\
&=&\sum_{q_1,q_2}\sum_{V_1,V_2}\sum_{pol}\int\limits_0^1 d\tau
\int\limits_{\frac{1}{2}\ln(\tau)}^
{-\frac{1}{2}\ln(\tau)} dy_q\,
f_{q_1}^{h_1}(\sqrt{\tau}e^{y_q},\mu_1^2)f_{q_2}^{h_2}(\sqrt{\tau}e^
{-y_q},\mu_2^2)
\nonumber\\
&&\cdot\int\limits_0^1d\hat{x}\,{\cal L}_{V_1,V_2,pol}^{q_1 q_2}
(\hat{x})\sigma(V_1 V_2\to \Xi,{\cal W}^2=\tau\hat{x}s_{hh}).
\label{drei}\end{eqnarray}

The expression (\ref{drei}) allows one to define luminosities
${\cal L}_{(V_1V_2)_{pol}}^{h_1h_2}(x)$
of vector-boson pairs in a hadron-pair,
\begin{equation}
\sigma(h_1 h_2 \to V_1 V_2\to \Xi, s_{hh})
=\sum_{(V_1,V_2)_{pol}}\int\limits_{x_{min}}^1 dx\,
{\cal L}_{(V_1V_2)_{pol}}^{h_1 h_2}(x)
\sigma(V_1 V_2\to \Xi,{\cal W}^2=x\,s_{hh}).
\label{sigpplum}\end{equation}
In (\ref{sigpplum}),
\begin{equation}
x\equiv{\cal W}^2/s_{hh}
\end{equation} 
is the ratio of the squares of the invariant masses of the
vector-boson pair and of the hadron-pair.
The minimum value for $x$ is given by $x_{min}=(M_1+M_2)^2/s_{hh}$.
The summation in (\ref{sigpplum}) extends over all (unordered)
vector-boson pairs $(V_1V_2)$ which can produce the state $\Xi$ and
the luminosities are given by the expression
\begin{eqnarray}
{\cal L}_{(V_1 V_2)_{pol}}^{h_1 h_2}(x)&=&
C_{(12)}\left(\frac{\alpha}{2\pi}\right)^2x\int\limits_0^{-\ln(x)}
\frac{d\left[\ln(\frac{1}{\tau})\right]}{\tau}
\left\{I_{y,pol}^{h_1 h_2}(\tau)
+I_{y,pol}^{h_2 h_1}(\tau)\right\}\cr
&&\cdot\int\limits_{\frac{1}{2}\ln(\hat{x})}
^{-\frac{1}{2}\ln(\hat{x})}d\hat{y}\,
{\cal L}_{pol}\left(\hat{x},\sqrt{\hat{x}}e^{\hat{y}},
\frac{M_1^2}{s_{qq}},\frac{M_2^2}{s_{qq}}\right),\label{pplumis_ex}
\end{eqnarray}
with
\begin{equation}
I_{y,pol}^{h_1 h_2}(\tau)
=\int\limits_{\frac{1}{2}\ln(\tau)}^
{-\frac{1}{2}\ln(\tau)}
dy_q\left(\sum_{q_1(V_1)}c_{q_1(V_1)}^{pol}f_{q_1}^{h_1}(\sqrt{\tau}
e^{y_q},\mu_1^2)\right)
\left(\sum_{q_2(V_2)}c_{q_2(V_2)}^{pol}f_{q_2}^{h_2}(\sqrt{\tau}e^{-{y_q}},
\mu_2^2)\right)\label{yint}.\end{equation}
The luminosities (\ref{pplumis_ex}) are exact in the sense that no
approximation has been made on the kinematics of the two vector-bosons.
We call (\ref{pplumis_ex}) with (\ref{dlumis_amp}) the exact luminosities.
In (\ref{pplumis_ex}), 
\begin{equation}
C_{(12)}\equiv\left\{\begin{array}{r@{\;;\quad}cl}
1&\mathrm{if}&V_1\neq V_2\\
1/2&\mathrm{if}&V_1=V_2\end{array}
\right.\label{Pairs}
\end{equation}
is a combinatorial factor.
We further introduced the variable 
\begin{equation}
\hat{y}\equiv\frac{1}{2}\ln\left(\hat{z}^2/\hat{x}\right)
=\frac{1}{2}\ln\left(\frac{\hat{z}_1}{\hat{z}_2}\right).
\end{equation} 
In the case of light-like momenta of the vector-bosons $V_1$ and $V_2$,
the variable $\hat{y}$ is the rapidity of the $(V_1 V_2)$ center-of-mass
motion in the quark-quark cms, taken along the
direction of motion of the quark from which $V_1$ was emitted.
The functions $I_{y,pol}^{h_1 h_2}(\tau)$ contain all dependence on the
type of the quarks $q_i(V_i)$, i.e., on the parton distribution functions
and on the quark couplings to the vector-bosons.
The remaining part of the $\tau$-integral in (\ref{pplumis_ex}) 
depends only on kinematical variables.
The summations in (\ref{yint}) extend over all quarks 
$q_1$ and $q_2$ which can couple to the vector-bosons $V_1$
and $V_2$, respectively.
In the derivation of (\ref{pplumis_ex}), (\ref{yint}) 
we made use of the symmetry property of the luminosities for vector-boson pairs
in fermion-pairs,
\begin{equation}
\int d\hat{y}\,{\cal L}_{pol}\left(\hat{x},\sqrt{\hat{x}}e^{\hat{y}},
\frac{M_1^2}{s_{qq}},\frac{M_2^2}{s_{qq}}\right)
=\int d\hat{y}\,{\cal L}_{\overline{pol}}\left(\hat{x},\sqrt{\hat{x}}e^{\hat{y}},
\frac{M_2^2}{s_{qq}},\frac{M_1^2}{s_{qq}}\right),
\end{equation}
where $\overline{pol}$ is obtained from $pol$ by exchanging the
helicities of $V_1$ and $V_2$ (i.e. $TL\to LT$, $TT\to TT$ etc.).

\begin{figure}
\begin{center}
\epsfig{file=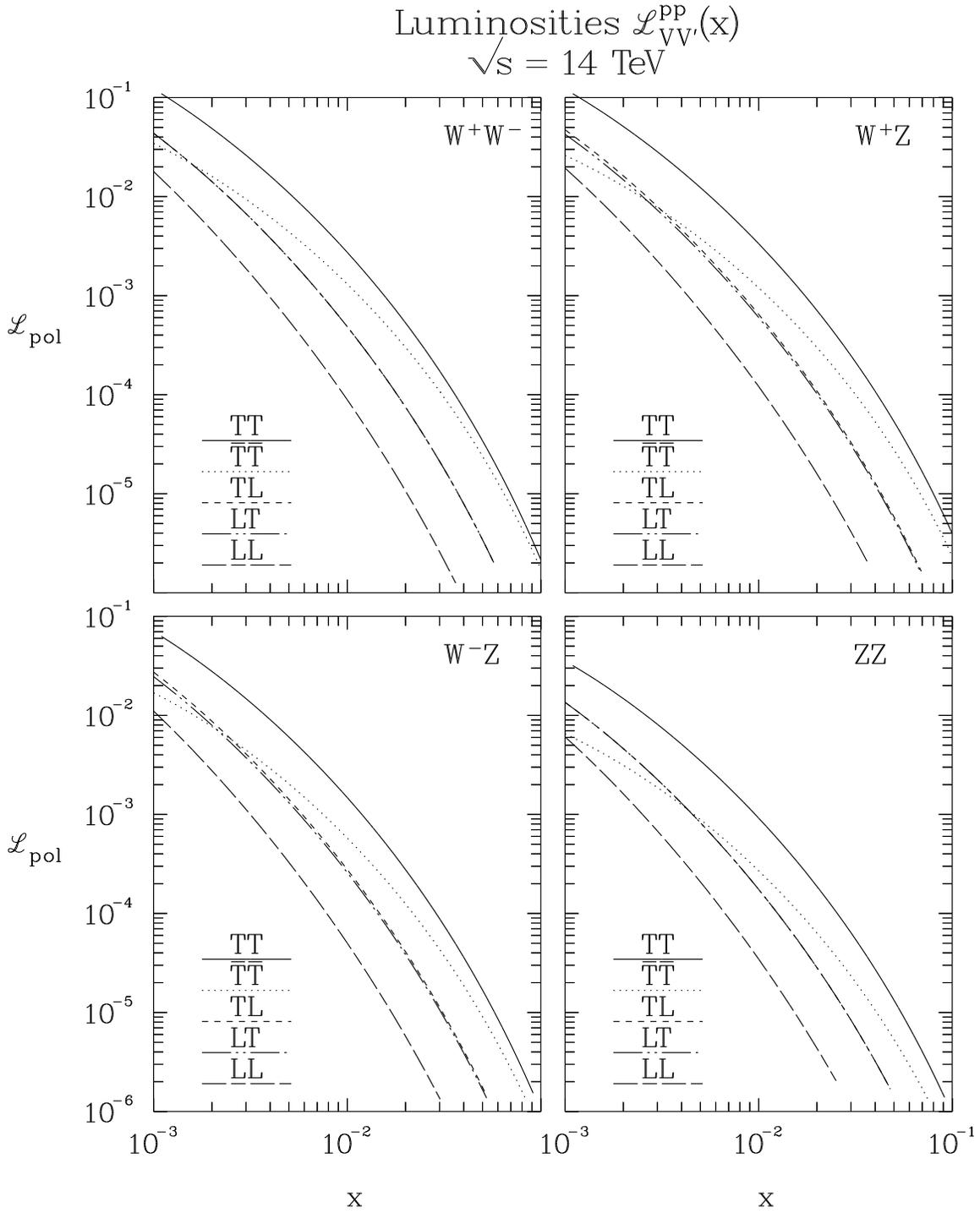,width=16cm,height=22cm}
\end{center}
\caption{The exact luminosities ${\protect\cal L}_{V_1V_2}^{h_1h_2}(x)$,
Eq. (\ref{pplumis_ex}) using (\ref{dlumis_amp}), 
for finding a vector-boson pair inside a proton-pair of $\protect\sqrt{s_{hh}}
=14$ TeV for the diagonal helicity combinations of
various vector-boson pairs as a function of the variable $x$.}
\label{lumx}
\end{figure}

Figure \ref{lumx} shows the exact luminosities (\ref{pplumis_ex}) 
with (\ref{dlumis_amp}) for the
vector-boson pairs $W^+W^-$, $W^+Z$, $W^-Z$ and $ZZ$ in a
proton-pair of $\sqrt{s}_{hh}=14$ TeV for the diagonal
helicity combinations as a function of $x$.
The definition for the helicity combinations $TT,\overline{TT},TL,LT$ and $LL$
can be found, e.g., in \cite{lumis}. 
The MRS(A) parametrization \cite{mrsa} in the DIS-scheme
was used for the parton distributions\footnote{
We use $\mu_i^2=\xi_is_{hh}$ unless explicitly stated otherwise. 
The sensitivity to the choice of the scale is small.} 
$f_{q_i}^p(\xi_i,\mu_i^2)$.
The electroweak parameters were $\alpha=1/128,M_W=80.17$ GeV and
$M_Z=91.19$ GeV. 

\subsection{Approximate Luminosities}
\begin{figure}[t]
\begin{center}
\epsfig{file=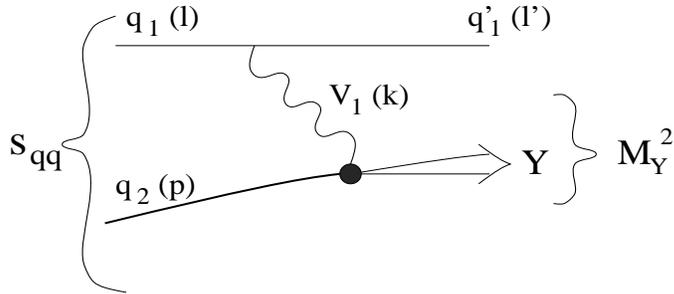,width=8.9cm,height=4cm}
\end{center}
\caption{The Feynman diagram in the effective vector-boson approximation
for the scattering of a quark $q_1$ with a
quark $q_2$ at a scattering energy $\protect\sqrt{s_{qq}}$. 
A final state $Y$ with the squared invariant mass 
$M_Y^2$ is produced. 
The particle $V_1$ is an exchanged vector-boson. The four-momenta
of the particles are denoted by $l,l',k$ and $p$.}
\label{qqdia}
\end{figure}
We give an approximation to the exact luminosities
which is obtained from
the full expression (\ref{dlumis_amp}).
It has been shown in \cite{lumis} how the expression (\ref{wichtig})
reduces to
a convolution of vector-boson distributions if certain kinematical
approximations are made.
The approximate expression for (\ref{wichtig}) is given by
\begin{equation}
{\cal L}_{V_1,V_2,\lambda_1\lambda_2}^{q_1q_2}(\hat{x})
=\eta_0\int\limits_{\hat{x}}^1\frac{d\hat{z}}{\hat{z}} 
{f}_{V_1,\lambda_1}^{q_1}\left(
\hat{z},\frac{M_1^2}{s_{qq}}\right)
{f}_{V_2,\lambda_2}^{q_2}\left(\frac{\hat{x}}{
\hat{z}},\frac{M_2^2}{\hat{z}\,s_{qq}}
\right),
\label{conv0}
\end{equation}
where 
the functions ${f}_{V_T}^{q}$, ${f}_{V_{\overline{T}}}^{q}$
and ${f}_{V_L}^{q}$ (we use $V=V_i$, $q=q_i$  etc. 
if no distinction between two different particles is necessary)
are the distribution functions of vector-bosons
in fermions of \cite{tung1}\footnote{No correction for the flux factor has to be
applied to the distributions \cite{tung1} appearing
in (\ref{conv0}) (as opposed to the prescription given
in Appendix \ref{appa}). This is because the boson-boson flux factor already
appears explicitly in front of the integral in (\ref{conv0}). It does
not have to be approximated as a product of boson-quark flux factors.}.
The label $\lambda=T,\overline{T},L$ denotes the helicity of the vector-boson.
Distribution functions of vector-bosons in fermions describe the 
process shown in Figure \ref{qqdia}. 
The cross-section for the process in Figure \ref{qqdia}, averaged
over the helicity of the quark $q_1$ and summed over the helicity of the
quark $q'_1$, is given in terms of $f_{V_\lambda}^q$ by
the expression
\begin{equation}
\sigma(q_1q_2\to q_1'Y,s_{qq})=\sum_{V_1,\lambda_1}\int\limits_{0}^1d\hat{z}
f_{V_{1,\lambda_1}}^{q_1}\left(\hat{z},\frac{M_1^2}{s_{qq}}\right)
\sigma(V_{1,\lambda_1}q_2\to Y,M_Y^2=\hat{z}s_{qq}).
\label{qqcross}\end{equation}
The distribution function $f_{V_{\lambda}}^q\left(\hat{z},
M^2/s_{qq}\right)$ is the probability density for the emission of a 
vector-boson $V$ with the helicity $\lambda$
and mass $M$ from a fermion $q$.
Separating the fermionic couplings, the $f_{V_\lambda}^q$ can be written as
\begin{equation}
f_{V_{\lambda}}^{q}\left(\hat{z},\frac{M^2}{s_{qq}}\right)
=\frac{\alpha}{2\pi}\hat{z}c_{q(V)}^{\lambda}
h_{\lambda}\left(\hat{z},\frac{M^2}{s_{qq}}\right).
\label{ampintro}\end{equation}
The quantities $h_{\lambda}$ in (\ref{ampintro}) 
are ``amputated'' vector-boson
distribution functions. The specific functions of \cite{tung1}
have to be taken. Amputated distribution
functions do not depend on the fermionic couplings
but only on two dimensionless variables.
Equivalent to (\ref{conv0}), for the corresponding
amputated differential luminosities one
obtains the forms
\begin{equation}
{\cal L}_{\lambda_1\lambda_2}\left(\hat{x},\hat{z},
\frac{M_1^2}{s_{qq}},\frac{M_2^2}{s_{qq}}\right)
=\eta_0\,{h}_{\lambda_1}\left(\hat{z},\frac{M_1^2}{s_{qq}}\right)
{h}_{\lambda_2}\left(\frac{\hat{x}}{\hat{z}},
\frac{M_2^2}{\hat{z}\,s_{qq}}\right).
\label{conv_asym}
\end{equation}
The amputated distribution functions ${h}_T$,
${h}_{\overline{T}}$ and ${h}_L$ 
have been given in closed form in \cite{tung1}.

We introduce a useful approximation to (\ref{conv_asym}).
The form of the luminosities (\ref{conv0}), (\ref{conv_asym}) 
is not invariant under the simultaneous exchange of the
vector-bosons $V_1$ and $V_2$ (i.e., their masses, fermionic couplings
and helicities $\lambda_1$ and $\lambda_2$) and the fermions $q_1$ and $q_2$.
The luminosities (\ref{wichtig}), however, obey this symmetry.
The symmetry is broken by the approximations which led to
(\ref{conv_asym}).
The symmetry of (\ref{conv_asym}) is not present
because different energies are available for the emission processes
of $V_1$ and $V_2$. The available energy for the emission of $V_2$
is reduced if $V_1$ is emitted in addition.
We extract from this fact
that an approximation to (\ref{wichtig}) should
feature an effective reduction of the available fermion-fermion scattering
energy. The reduction is due to the simultaneous emission of two vector-bosons.
We introduce the symmetrized form
\begin{equation}
{\cal L}_{\lambda_1\lambda_2}\left(\hat{x},\hat{z},
\frac{M_1^2}{s_{qq}},\frac{M_2^2}{s_{qq}}\right)=
\eta_0\,{h}_{\lambda_1}\left(\hat{z},\frac{M_1^2}{\sqrt{\hat{z}}s_{qq}}\right)
{h}_{\lambda_2}\left(\frac{\hat{x}}{\hat{z}},
\frac{M_2^2\,\sqrt{\hat{z}}}{\sqrt{\hat{x}}s_{qq}}\right).
\label{convsym}
\end{equation}
In (\ref{convsym}),
the available energies for the emission of $V_1$ and $V_2$
are $\sqrt{\hat{z}}s_{qq}$ and $\sqrt{\hat{x}/\hat{z}}s_{qq}$, respectively.
Eq. (\ref{convsym}) is an approximation to Eq. (\ref{conv_asym}).
We refer to the luminosities (\ref{pplumis_ex}) using (\ref{convsym})
as Approximation 1.

\begin{figure}
\begin{center}
\epsfig{file=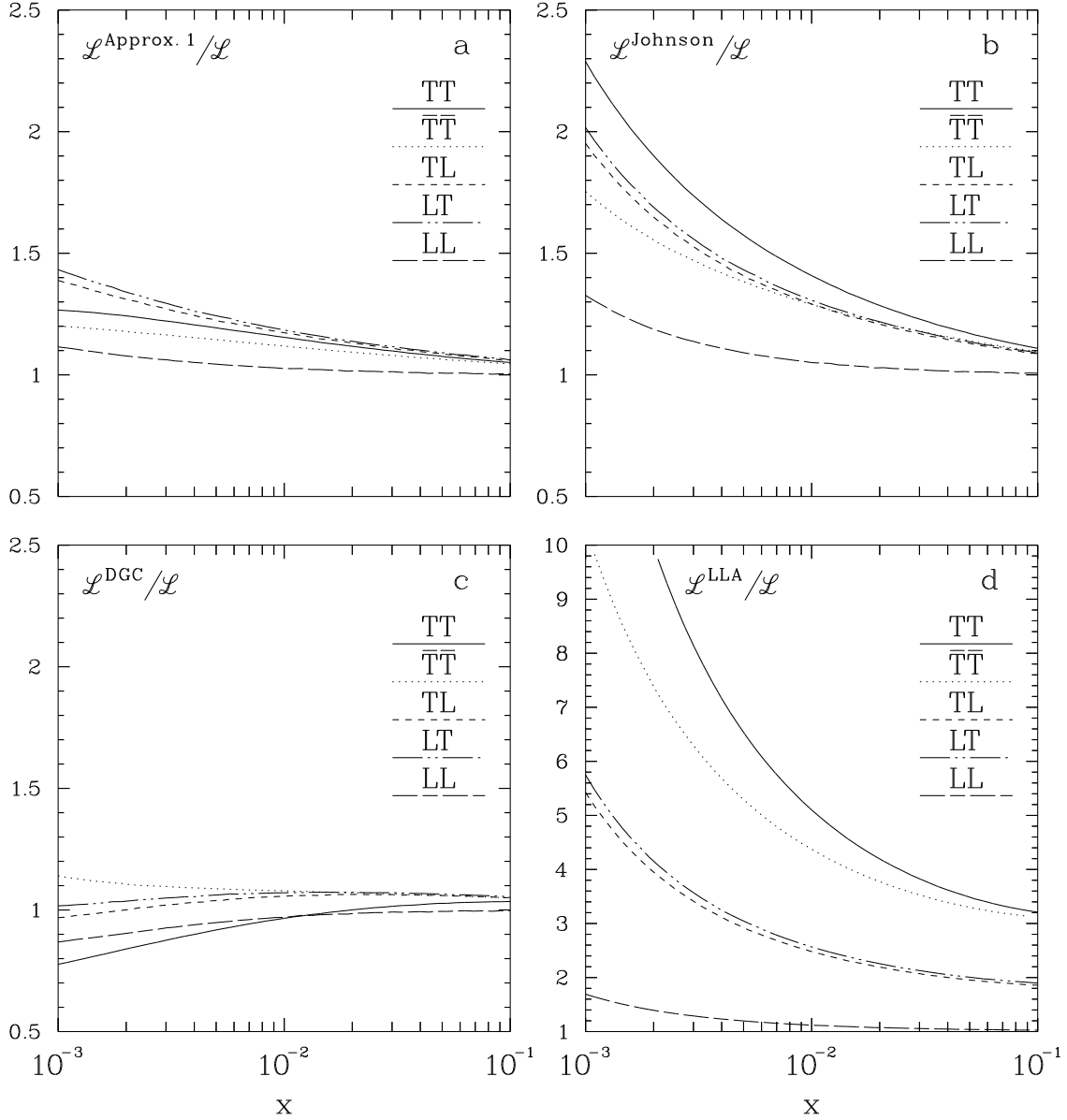,width=15cm,height=20.5cm}
\end{center}
\caption{The ratio of the luminosities using the approximations (\ref{convsym})
or (\ref{hprod})
and the exact luminosities, Eq. (\ref{dlumis_amp}), 
for finding a $W^+Z$ pair in a proton pair of
$\protect\sqrt{s_{hh}}=14$ TeV for the diagonal helicity combinations
as a function of the variable $x$.
In Figure a, the direct approximation,
Eq. (\protect\ref{convsym}), was used.
In Figures b, c and d, the vector-boson distributions of \protect\cite{tung1},
Eq. (\ref{integrals}) (DGC) and the LLA, respectively,
were used to evaluate 
${\protect\cal L}_{pol}$ according to Eq. (\protect\ref{hprod}). 
All luminosities have been
calculated according to Eq. (\ref{pplumis_ex}).}
\label{ratio1}
\end{figure}
Figure \ref{ratio1} a shows the ratios of the luminosities (\ref{pplumis_ex}),
evaluated with (\ref{convsym}), and the exact luminosities,
(\ref{pplumis_ex}) with (\ref{dlumis_amp}), for the diagonal helicity 
combinations as a function of $x$. 
The region $10^{-3}\lsim x\lsim 2\cdot 10^{-2}$ is particularly
interesting for vector-boson pair production. It corresponds to
invariant masses of $400$ GeV $\lsim M_{V_3V_4}\lsim 2$ TeV. In this
region, the dominant
$TT$ luminosity deviates by less than $30\%$ from the exact result.
A comparison with results of other authors will be given in Section \ref{2.3}.

Further approximations may be applied.
So far we have employed approximate expressions for the
${\cal L}_{V_1,V_2,pol}^{q_1q_2}(\hat{x})$ but we used
the correct expression for the quark-quark
subenergy, $s_{qq}\equiv\tau s_{hh}$. If approximate expressions for $s_{qq}$
are used,
the luminosity ${\cal L}_{(V_1V_2)_{pol}}^{h_1h_2}(x)$ can be approximated as
a convolution of vector-boson distribution functions $f_{V_{\lambda}}^h$
in hadrons. 
Luminosities have been previously obtained in this way
\cite{lindfors,capdequi}.  
The possibility to use vector-boson distributions
in hadrons has already been mentioned in \cite{dawson}.
An equivalent expression for (\ref{pplumis_ex}) is
\begin{eqnarray}
{\cal L}_{(V_1 V_2)_{pol}}^{h_1 h_2}(x)&=&C_{(12)}
\left(\frac{\alpha}{2\pi}\right)^2
x\int\limits_x^1\frac{dz}{z}\int\limits_{z}^1\frac{d\xi_1}{\xi_1}
\int\limits_{x/z}^1\frac{d\xi_2}{\xi_2}\frac{1}{\xi_1\xi_2}
\nonumber\\
&&\times\left[\left(\sum_{q_1(V_1)}c_{q_1(V_1)}^{pol} f_{q_1}^{h_1}(\xi_1,\mu_1^2)
\right)\cdot
\left(\sum_{q_2(V_2)}c_{q_2(V_2)}^{pol} f_{q_2}^{h_2}(\xi_2,\mu_2^2)\right)
\;\;+\;\; h_1\leftrightarrow h_2\,\right]\cr
&&\cdot{\cal L}_{pol}\left(\hat{x}=\frac{x}{\xi_1\xi_2},\hat{z}=\frac{z}{\xi_1},
\frac{M_1^2}{s_{qq}},\frac{M_2^2}{s_{qq}}\right),
\label{pplumis_equiv}\end{eqnarray}
where we introduced the variable $z=\xi_1\hat{z_1}$.
The variable $z$
describes the invariant mass squared 
$M_Y^2$
which is left for the reaction of the vector-boson $V_1$ with
the quark $q_2$.
If $V_1$ is light-like, $z$ is equal to 
the ratio of the energy of $V_1$ and the
energy of the hadron from which it was emitted. 
In analogy to $\hat{z}_1$ and $\hat{z}_2$ we introduce the variables
$z_1=z$ and $z_2=x/z$.
It follows that $z_i=\xi_i\hat{z}_i$.
Inserting the convolutions (\ref{convsym}) into (\ref{pplumis_equiv})
leads to the expression
\begin{eqnarray}
{\cal L}_{(V_1 V_2)_{\lambda_1\lambda_2}}^{h_1 h_2}(x)
&=&C_{(12)}\,\eta_0\int\limits_x^1\frac{dz}{z}
\int\limits_{z}^1d\xi_1
\int\limits_{x/z}^1d\xi_2
\nonumber\\
&&\cdot\left[\frac{df_{V_{1,\lambda_1}}^{h_1}}{d\xi_1}(z)
\frac{df_{V_{2,\lambda_2}}^{h_2}}{d\xi_2}\left(\frac{x}{z}\right)
\;\;+\;\; h_1\leftrightarrow h_2\,\right].\label{nofactor}
\end{eqnarray}
In (\ref{nofactor}), the quantities
$df_{V_{\lambda}}^{h}/d\xi$ are
differential distribution-functions of a vector-boson
$V_{\lambda}$ in a hadron $h$. They are given by
\begin{eqnarray}
\xi\frac{df_{V_{\lambda}}^{h}}{d\xi}(z)
&=&\frac{\alpha}{2\pi}\frac{z}{\xi}\sum_{q(V)} c_{q(V)}^{\lambda}
f_{q}^{h}(\xi,\mu^2)h_{\lambda}\left(\frac{z}{\xi},
\frac{M^2\sqrt{\xi}}{\sqrt{z}\,s_{qq}}\right),
\label{diffvecinps}\end{eqnarray}
with $s_{qq}=\xi_1\xi_2 s_{hh}$.

The integrations over $\xi_1$ and $\xi_2$ in (\ref{nofactor}) cannot be
carried out independently because $s_{qq}$
in the differential distributions (\ref{diffvecinps}) depends on
both $\xi_1$ and $\xi_2$. We may, however,
approximate $s_{qq}$ by $s_{qq}=\xi_i^2
s_{hh}$. It means that we assume the same energy, $E=\xi_iE_h$, for the quark
which emits the vector-boson $V_i$ and the other quark which emits $V_j$.
$E_h$ is the hadron energy evaluated in the hadron-hadron cms,
$E_h=\sqrt{s_{hh}}/2$.
Equivalently, it means that the parton-parton cms is approximated as the
hadron-hadron cms. 
With this approximation the vector-boson distributions in a hadron are
given by
\begin{equation}
f_{V_{\lambda}}^{h}(z)=\int\limits_{z}^1\frac{d\xi}{\xi}
\sum_{q(V)}f_{q}^{h}(\xi,\mu^2)
f_{V_{\lambda}}^{q}\left(\frac{z}{\xi}
,\frac{M^2\sqrt{\xi}}{\sqrt{z}s_{qq}}\right),
\label{vecinp2}\end{equation}
with $s_{qq}=\xi^2s_{hh}$ and $f_{V_{\lambda}}^{q}(\hat{z}=z/\xi)$
from (\ref{ampintro}). We require
$z/\xi>M^2/s_{qq}$, i.e., $\xi_{min}=\max[z,M^2/(z s_{hh})]$ as the lower
limit of integration in (\ref{vecinp2}).
Again, as in (\ref{conv0}), 
the functions (\ref{vecinp2}) do not contain a flux
factor since the boson-boson flux factor $\eta_0$ already
appears explicitly in front of the integral in
(\ref{nofactor}). 

The luminosities ${\cal L}_{(V_1 V_2)}^{h_1 h_2}(x)$ are
given by using the functions (\ref{vecinp2}) in Eq.
(\ref{nofactor}),
\begin{equation}
{\cal L}_{(V_1 V_2)_{\lambda_1\lambda_2}}^{h_1 h_2}(x)
=C_{(12)}\,\eta_0\int\limits_{\frac{1}{2}\ln(x)
}^{-\frac{1}{2}\ln(x)}dy
\left[f_{V_{1,\lambda_1}}^{h_1}\left(\sqrt{x}e^{y}\right)
f_{V_{2,\lambda_2}}^{h_2}\left(\sqrt{x}e^{-y}\right)
\;+\; h_1\leftrightarrow h_2\,\right],\label{lumiconv2}
\end{equation}
where we introduced the variable
\begin{equation}
y\equiv \frac{1}{2}\ln\left(z^2/x\right)
=\frac{1}{2}\ln\left(\frac{z_1}{z_2}\right)=y_q+\hat{y}.
\end{equation} 
If the vector-bosons $V_1,V_2$ are light-like, $y$
is the rapidity of the $V_1V_2$ center-of-mass motion taken along
the direction of motion of the hadron which emitted $V_1$.
The formula (\ref{lumiconv2}) 
has been derived with only the mentioned approximations
(using the factorized forms (\ref{convsym}) and approximating $s_{qq}$
by $s_{qq}\simeq \xi^2 s_{hh}$)
from the
exact luminosities for a vector-boson pair in a proton-pair,
(\ref{pplumis_ex}) with (\ref{dlumis_amp}).
We refer to (\ref{lumiconv2}) as Approximation 2.
It is a direct approximation to the exact luminosities.

\begin{figure}
\begin{center}
\epsfig{file=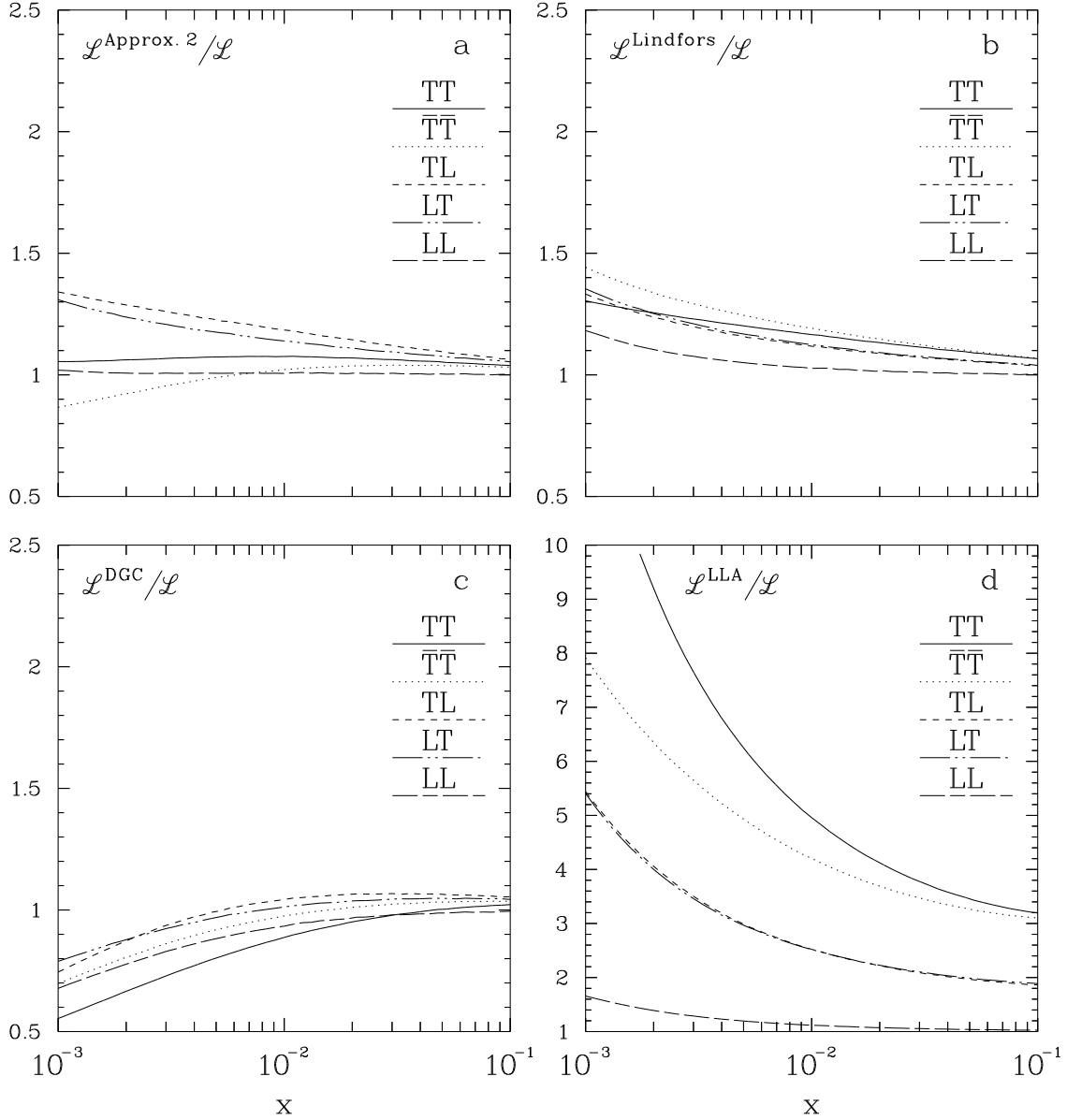,width=15cm,height=20.5cm}
\end{center}
\caption{The ratio of the luminosities approximated as convolutions
of vector-boson distribution functions $f_{V_{\lambda}}^p$
and the exact luminosities
for finding a $W^+Z$ pair in a proton pair of 
$\protect\sqrt{s_{hh}}=14$ TeV for the diagonal helicity combinations 
as a function of the variable $x$. 
In Figure a, Eq. (\ref{lumiconv2})
was used to evaluate
$f_{V_{\lambda}}^q$ in (\ref{vecinp}).
In Figures b, c and d, Eq. (\ref{lumiconv}) was used and
the
vector-boson distributions of \protect\cite{lindfors},
Eq. (\ref{integrals}) (DGC) and the LLA, respectively, were used.}
\label{ratio2}
\end{figure}
Figure \ref{ratio2} a shows the ratios of the luminosities in
Approximation 2,
Eq. (\ref{lumiconv2}), and the exact luminosities
for finding a $W^+Z$ pair
in a proton pair of $\sqrt{s_{hh}}=14$ TeV 
for the diagonal helicity combinations as a function of $x$.
The Approximation 2 is in excellent agreement with the improved EVBA.
The results of other authors will be discussed in the following 
Section.

\subsection{Comparison with the Literature\label{2.3}}
The exact luminosities (\ref{pplumis_ex}) with (\ref{dlumis_amp})
may be compared with 
results presented in the literature \cite{dawson,lindfors,capdequi}.
In contrast to the approximations (\ref{convsym}) and (\ref{lumiconv2}),
these results do not use the exact expression as a starting point. 
Instead, they use the ad hoc
assumption that the luminosities ${\cal L}_{V_1,V_2,pol}^{q_1q_2}(\hat{x})$
can be
obtained by 
convolutions of vector-boson distribution functions.
The convolution is similar to (\ref{conv0}) and is
given by
\begin{equation}
{\cal L}_{V_1,V_2,\lambda_1\lambda_2}^{q_1q_2}(\hat{x})
=\int\limits_{\hat{x}}^1\frac{d\hat{z}}{\hat{z}}
f_{V_{1,\lambda_1}}^{q_1}\left(\hat{z},\frac{M_1^2}{s_{qq}}\right) 
f_{V_{2,\lambda_2}}^{q_2}\left(\frac{\hat{x}}{\hat{z}},
\frac{M_2^2}{s_{qq}}\right).
\label{qqconv}\end{equation}
Instead of the particular 
functions $f_{V_{\lambda}}^q$ of \cite{tung1}, various different
functions have been used.
Equivalent to the approximation
(\ref{qqconv}), the amputated differential 
luminosities are written as a product of amputated distribution
functions,
\begin{equation}
{\cal L}_{\lambda_1\lambda_2}
\left(\hat{x},\hat{z},\frac{M_1^2}{s_{qq}},\frac{M_2^2}{s_{qq}}
\right)=h_{\lambda_1}\left(\hat{z},\frac{M_1^2}{s_{qq}}\right)
h_{\lambda_2}\left(\frac{\hat{x}}{\hat{z}},\frac{M_2^2}{s_{qq}}\right).
\label{hprod}\end{equation}

Vector-boson distribution functions have been derived by several
authors
\cite{dawson,karero,lindfors,capdequi,tung1,ardv,rolnick,godbole}.
In most cases more assumptions than
only the ones inherent in the EVBA, i.e. that the reaction
proceeds via the exchange of vector-bosons and that the vector-boson
scattering cross-sections for off-shell vector-bosons must be known
(or an assumption has to be made), were made in the derivation. 
The differences of
various derivations are discussed in Appendix \ref{appa}.
In the Appendix we also specify
the vector-boson distribution functions $f_{V_\lambda}^q$ 
which we use for our
numerical examples.

Figures \ref{ratio1} b,c and d show the 
ratios of the approximated luminosities, 
Eq. (\ref{pplumis_ex}) evaluated 
with (\ref{hprod}), and the exact luminosities, (\ref{pplumis_ex})
with (\ref{dlumis_amp}), using for $h_{\lambda}$ the distributons
\cite{tung1}, the distributions (\ref{integrals}) and the LLA\footnote{
We use $s_{qq}/M^2$ as the arguments of the logarithms. This is
the simplest choice. Other choices and next-to-leading
forms have been used e.g. in \cite{lindfors2}.}, respectively,
for the diagonal helicity combinations 
as a function of $x$. 
The LLA overestimates the improved EVBA by an order of magnitude at small
$x$ if both polarizations are transverse.
Using (\ref{integrals}) 
or \cite{tung1}, instead,
greatly diminishes the deviation of the approximation from the improved EVBA.
We note that the better agreement 
of the distributions (\ref{integrals}) than the one of the
distributions \cite{tung1}
with the improved EVBA is accidental since 
the distributions (\ref{integrals}) involve
additional approximations as compared to the distributions \cite{tung1}
(see Appendix \ref{appa}).
One sees that 
the use of Eq. (\ref{convsym}) (Figure \ref{ratio1} a) further improves the
agreement between approximated and exact luminsoities, 
at least compared to the convolutions of \cite{tung1}.
In particular,
the agreement is substantially improved for the dominating 
$TT$ luminosity.

\begin{figure}
\begin{center}
\epsfig{file=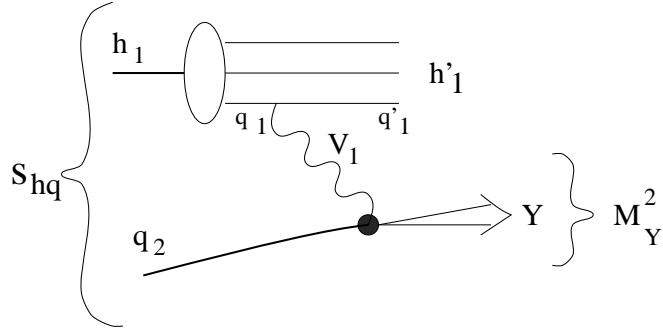,width=8.7cm,height=4.4cm}
\end{center}
\caption{The diagram for the scattering of a hadron $h_1$ and
a quark $q_2$ proceeding via the exchange of a single vector-boson $V_1$
originating from the hadron.
A final state $Y$ of invariant mass $M_Y$ is produced.}
\label{fvhdia}
\end{figure}
Similarly to Approximation 2, 
vector-boson distribution functions in hadrons have been 
used. They were derived in order to describe the process shown 
in Figure \ref{fvhdia}.
The distribution functions were
obtained as convolutions
of the quark distributions in hadrons and the vector-boson distributions
in quarks,
\begin{equation}
f_{V_{\lambda}}^{h}(z)=\int\limits_{z}^1\frac{d\xi}{\xi}
\sum_{q(V)}f_{q}^{h}(\xi,\mu_i^2)
f_{V_{\lambda}}^{q}\left(\frac{z}{\xi},\frac{M^2}{s_{qq}}\right).
\label{vecinp}\end{equation}
The functions $f_{V_{\lambda}}^{h}$ describe the emission probability of
a vector-boson $V$ with helicity $\lambda$ and mass $M$ 
from a hadron $h$.
The sum in (\ref{vecinp}) extends over all quarks and antiquarks which
can couple to $V$. Eq. (\ref{vecinp}) is similar to (\ref{vecinp2}).
However, in (\ref{vecinp2}) the specific distributions \cite{tung1}
have to be used. In addition, in (\ref{vecinp2})
we included the square roots introduced in (\ref{convsym}).
The definition of $z$ in terms of $s_{hq}$ (defined in Figure \ref{fvhdia})
and $M_Y^2$ is given by
\begin{equation}
z\equiv M_Y^2/s_{hq}.
\label{zdef}\end{equation}
The cross-section for the process shown in Figure \ref{fvhdia} is given
by
\begin{equation}
\sigma(h_1q_2\to h_1'Y,s_{hq})=\sum_{V_1,\lambda_1}\int\limits_0^1 dz
f_{V_{1,\lambda_1}}^{h_1}\left(z,\frac{M_1^2}{s_{qq}}\right)
\sigma(V_{1,\lambda_1}q_2\to Y,M_Y^2).\label{hqcross}
\end{equation}
In writing down (\ref{hqcross}) no other assumptions than those inherent
in the EVBA have been made.

The quark-quark energy $s_{qq}$ is in principle
unknown if only the energy $E_h$ of the hadron or the
hadron-hadron scattering energy $s_{hh}$ is known. Thus,
an approximation for $s_{qq}$ has to be made. We will again
use $s_{qq}\simeq \xi^2 s_{hh}$, as above.
 
In \cite{lindfors2}, a variable $Q^2$, which was defined
by $Q^2\equiv M_Y^2$ was used. 
The approximation $s_{qq}\simeq \xi^2 s_{hh}$ applied to $Q^2$ is
$Q^2=\hat{z}s_{qq}=z/\xi s_{qq}\simeq z\xi s_{hh}$.

The luminosities ${\cal L}_{(V_1V_2)_{\lambda_1\lambda_2}}^{h_1h_2}(x)$
were approximately expressed as convolutions of the vector-boson
distribution functions (\ref{vecinp}),
\begin{equation}
{\cal L}_{(V_1 V_2)_{\lambda_1\lambda_2}}^{h_1 h_2}(x)
=C_{(12)}\int\limits_{\frac{1}{2}\ln(x)
}^{-\frac{1}{2}\ln(x)}dy
\left[f_{V_{1,\lambda_1}}^{h_1}\left(\sqrt{x}e^{y}\right)
f_{V_{2,\lambda_2}}^{h_2}\left(\sqrt{x}e^{-y}\right)
\;+\; h_1\leftrightarrow h_2\,\right].\label{lumiconv}
\end{equation}
The approximation (\ref{lumiconv}) has often been used in the past 
and numerical results for the luminosities can be found in 
\cite{capdequi,lindfors2}. 

We note that in \cite{lindfors2} an excellent approximation was made.
As discussed above,
instead of the variable $s_{qq}$ two
variables $Q_1^2$ and $Q_2^2$, which appear in $f_{V_{1,\lambda_1}}^{h_1}$
and $f_{V_{2,\lambda_2}}^{h_2}$, respectively,
were used.  
The $Q_i^2$ are the squared invariant masses of a vector-boson
and a quark which are
defined in terms of $s_{qq}$, or, alternatively, in terms of $s_{hh}$
by
\begin{equation}
Q_i^2\equiv \hat{z}_is_{qq}=\frac{z_i}{\xi_i} s_{qq}=x\frac{\xi_j}{z_j}s_{hh},
\quad i\neq j,\quad i=1,2,
\label{q2def}\end{equation}
where we have only used the exact relations
$s_{qq}=\xi_1\xi_2s_{hh}$ and $z_1z_2=x$. 
Clearly, again, factorization does not occur
using the exact expressions (\ref{q2def}). 
Instead of using the approximation
$s_{qq}\simeq \xi_i^2s_{hh}$, another approximation
for the $Q_i^2$ has been made in \cite{lindfors2} when
luminosities were calculated, namely the simple approximate
choice $Q_i^2\simeq xs_{hh}={\cal W}^2$.
Thus, the quark vector-boson invariant masses $Q_i^2$ have
been approximated by the vector-boson vector-boson invariant mass.
This choice always underestimates\footnote{
We see from (\ref{q2def}) that the approximated values for the
$Q_i^2$ are smaller by the factors
$z_j/\xi_j<1$ than the exact values. One should note that the variable
$\xi_i$ runs in the limits $z_i<\xi_i<1$.} the exact values of $Q_i^2$.
However, we know that reducing the invariant masses involving quarks
will in general improve the agreement with the improved EVBA.
We will therefore use this choice of $Q_i^2$ in our numerical example below.
It leads to an excellent agreement with the improved EVBA.
For the other distributions $f_{V_{\lambda}}^h$, 
we use $s_{qq}\simeq \xi_i^2s_{hh}$ as
before.

Figures \ref{ratio2} b,c and d show the ratios of the luminosities 
calculated according to Eq. (\ref{lumiconv})
and the exact luminosities
for finding a $W^+Z$ pair
in a proton pair of $\sqrt{s_{hh}}=14$ TeV 
for the diagonal helicity combinations as a function of $x$.
To evaluate (\ref{lumiconv}), the distributions
of \cite{lindfors} (using $Q_i^2={\cal W}^2$), the distributions
(\ref{integrals}) and the LLA
were used. The LLA overestimates the exact luminosities by far.
The distributions (\ref{integrals}) yield slightly low values at
low $x$. The distributions \cite{lindfors} 
are an excellent approximation to the improved EVBA. 
For the dominant $TT$ luminosity, the direct approximation (Figure
\ref{ratio2} a) is better than the distributions \cite{lindfors}.

\begin{figure}
\begin{center}
\epsfig{file=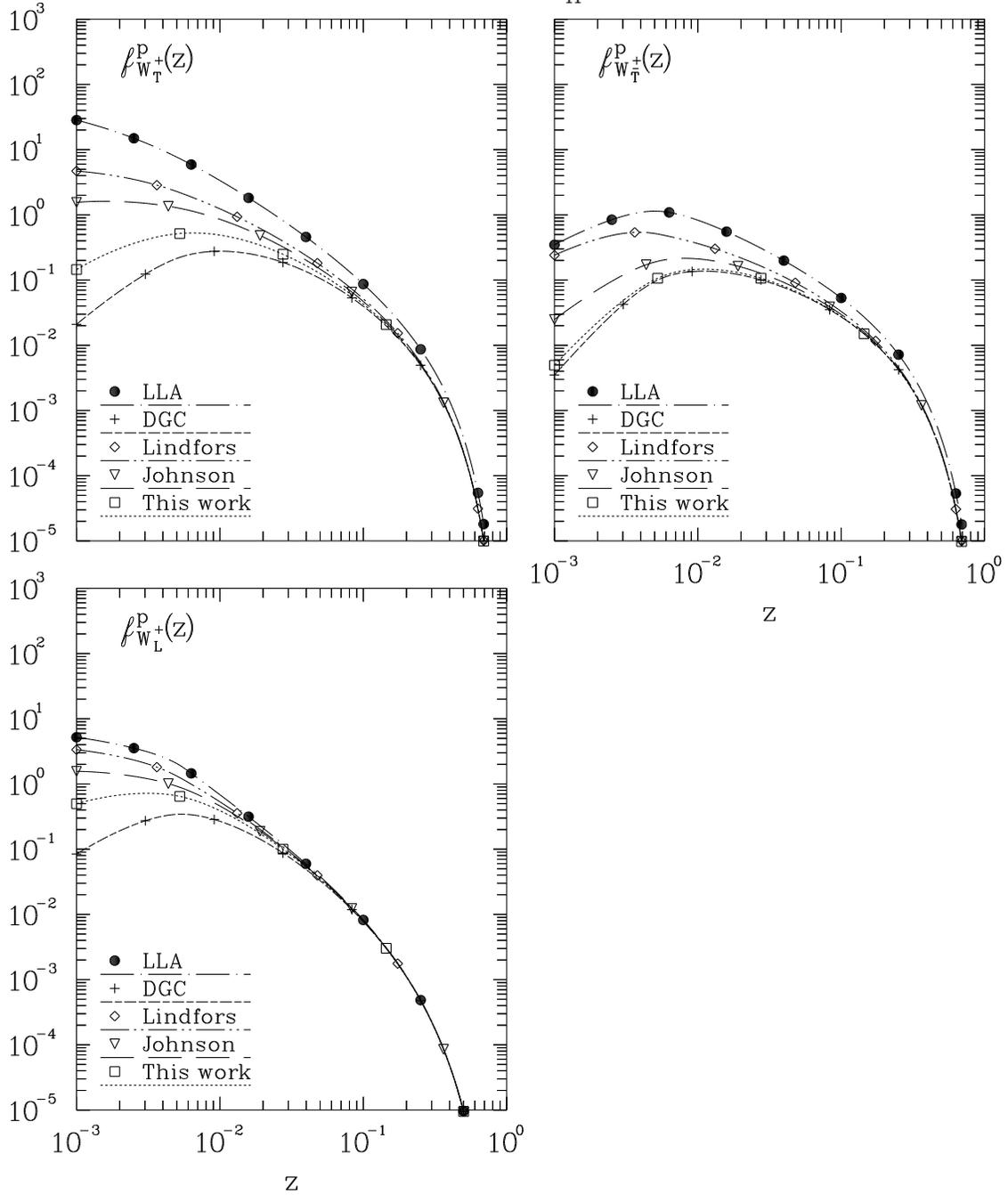,width=15cm,height=21cm}
\end{center}
\caption{The distribution functions of a $W^+$ boson in a proton of
$E_h=\protect\sqrt{s_{hh}}=7$ 
TeV, Eq. (\ref{vecinp2}) (This work) or (\ref{vecinp}) (all others),
for the helicity combinations $T$, $\protect\overline{T}$ and $L$ 
as a function of $z$. The LLA, Eq. (\ref{integrals}) (DGC)
and the distributions
\protect\cite{lindfors} and \protect\cite{tung1} were used to evaluate
$f_{V_{\lambda}}^q$ in (\ref{vecinp}).}
\label{fvpnum}
\end{figure}
We finally present numerical results for the vector-boson distribution
functions in ha\-drons.
In the numerical example of the functions (\ref{vecinp2}) 
we approximate the
boson-boson flux factor in (\ref{nofactor})
by a product of boson-quark flux factors,
\begin{equation}
\eta_0\simeq \left(1-\frac{M_1^2}{\hat{z}_1s_{qq}}\right)
\left(1-\frac{M_2^2}{\hat{z}_2s_{qq}}\right).
\end{equation}
One of the boson-quark flux factors, 
$\left(1-\frac{M^2}{\hat{z} s_{qq}}\right)$,
is then included in the 
$f_{V_{\lambda}}^{q}(\hat{z}=z/\xi)$ in (\ref{vecinp2}).
Figure \ref{fvpnum} shows the distributions functions for a $W^+$ 
in a proton of $E_h=\sqrt{s_{hh}}/2=7$ TeV for
the various helicity combinations of the $W^+$ as a function of $z$.
The distribution functions have been calculated according to Eq. 
(\ref{vecinp2}) or Eq. (\ref{vecinp}). To evaluate Eq. (\ref{vecinp}),
the LLA, the distributions (\ref{integrals}) and those of
\cite{lindfors} and \cite{tung1}
have been used for $f_{V_\lambda}^q$. 
We note that we used the complete
expressions \cite{lindfors} instead of the next-to leading forms
\cite{lindfors2}. 

For the $T$- and $\overline{T}$-polarization, 
the LLA overestimates any of the other distributions 
by far.
The distributions \cite{tung1} 
are larger than those of Eq. (\ref{vecinp2}). 
The distributions (\ref{integrals}) yield rather low values.
The differences between the distributions increase
at small $z$. 
For the $L$-polarization, the differences between the models only manifest
themselves at low values of $z$.

We note that a typical value for $z$ is $z=\sqrt{x}\simeq 7\cdot 10^{-2}$
if a final state $\Xi$ of mass 1 TeV is produced in $pp$-collisions
at $\sqrt{s_{hh}}=14$ TeV. All values of $z$ in the range $x<z<1$,
however, contribute to the integral in (\ref{lumiconv2}) or (\ref{lumiconv}).
For ${\cal W}=450$ GeV, which is still a large energy compared
to the vector-boson masses, $z$ becomes as small as $z=x\simeq 10^{-3}$
and the whole range of $z$ which is shown in Figure \ref{fvpnum} contributes
to the luminosity.

\section{Comparison with a Complete Perturbative Calculation\label{compsec}}
We are now going to present a numerical comparison of
the EVBA with a complete perturbative calculation.
The complete perturbative calculation includes the contribution from
bremsstrahlung diagrams as shown in Figure \ref{fig3}.
As an example for a vector-boson pair production process, 
we choose the process $pp\to ZZX$, for which complete results
are available in the literature.

The complete perturbative calculation uses in
Eq. (\ref{QPM2}) the complete (lowest
order) cross-section 
of the process on the quark-level, $q_1 q_2\to q_1'q_2' ZZ$.
Numerical results of the complete calculation for $\sqrt{s}=40$ TeV
can be found 
in \cite{dicus_vega} and \cite{ardv,abrep}.
Only the production via $W^+W^-$-pairs, 
$pp\to W^+W^-\to ZZ$, was considered\footnote{A separation
into a contribution from intermediate $W^+W^-$-pairs and a contribution
from intermediate $ZZ$-pairs is also possible in the complete calculation
(in a very good approximation) \cite{dicus_vega}.
The diagrams of the complete calculation can be grouped into two
classes. One class contains the $W^+W^-$-diagrams of the EVBA
and additional bremsstrahlung-type diagrams, the other class
contains the $ZZ$-diagrams and also additional bremsstrahlung-diagrams.
Both classes are a gauge-invariant subset. The interference term
between the two classes, which arises when the
amplitude is squared, is very small.}.

In the EVBA one has to calculate the cross-sections for
$W^+W^-\to ZZ$. 
In the Born approximation there are four diagrams which contribute to
these processes. Two diagrams describe the exchange of massive vector-bosons,
one diagram a four-particle point interaction and one diagram Higgs boson
exchange. An analytical expression for the helicity amplitudes
has been given in \cite{dipl}.

As in \cite{ardv}, we apply a rapidity cut on the 
produced vector-bosons $V_3V_4=ZZ$ in the 
hadron-hadron cms frame. We treat this cut approximately assuming that
the vector-bosons $V_1,V_2$ move collinearly to the hadron beam
direction. The rapidities of the vector-bosons $V_3$ and $V_4$ in the
$V_1V_2$ cms frame, taken along the direction of motion of the hadron
from which $V_1$ is emitted, are
\begin{equation}
y_3^\ast=\tanh^{-1}\left(\frac{q\cos\theta}{\sqrt{q^2+M_3^2}}\right),
\quad\quad y_4^\ast=\tanh^{-1}
\left(\frac{-q\cos\theta}{\sqrt{q^2+M_4^2}}\right).
\label{y3y4}\end{equation}
In (\ref{y3y4}), $\theta$ is the angle between the directions of motion
of $V_1$ and $V_3$ evaluated in the $(V_1V_2)$ center-of-mass system.
The variable
$q$ is the magnitude of the space-like momentum of the vector-boson $V_3$
in this system,
\begin{equation}
q=\frac{\sqrt{{\cal W}^2}}
{2}\sqrt{1-\frac{2}{{\cal W}^2}(M_3^2+M_4^2)+\frac{1}{{\cal W}^4}
(M_3^2-M_4^2)^2}.
\end{equation}
The rapidities $y_3,y_4$ of $V_3,V_4$ in the hadron-hadron cms frame
are approximately obtained by addition, 
$y_3\simeq y+y_3^\ast$ and $y_4\simeq y+y_4^\ast$, where
the equality holds strictly if both $V_1$ and $V_2$ are light-like.
We apply a rapidity cut $Y$ to both produced vector-bosons,
\begin{equation}
|y_3|<Y\;\;\mbox{and}\;\;|y_4|<Y.
\label{Ycutdef}\end{equation}
Following from (\ref{sigpplum}) and
(\ref{pplumis_ex}) with (\ref{yint}), we obtain the expression for the
cross-section for $h_1h_2\to V_1 V_2\to V_3 V_4$ with a rapidity cut,
\begin{eqnarray}
&&\frac{d\sigma}{dx}(h_1 h_2\to V_1 V_2\to V_3 V_4, s_{hh})
\theta(Y-|y_3|)\theta(Y-|y_4|)\cr
&=&\left(\frac{\alpha}{2\pi}\right)^2x
\int\limits_{-y_{\mathrm{max}}}^{y_{\mathrm{max}}}dy
\int\limits_{0}^{\ln\left(\frac{1}{x}\right)}\frac{d\ln\left(
\frac{1}{\tau}\right)}{\tau}
\int\limits_{\max[-\frac{1}{2}\ln\left(\frac{1}{\tau}\right),
-\frac{1}{2}\ln\left(\frac{\tau}{x}\right)+y]}^
{\min[\frac{1}{2}\ln\left(\frac{1}{\tau}\right),
\frac{1}{2}\ln\left(\frac{\tau}{x}\right)+y]}dy_q
\sum_{(V_1 V_2)}C_{(12)}\,\eta_0\sum_{pol}\cr
&&\cdot\Bigg[\left(\sum_{q_1(V_1)}f_{q_1}^{h_1}
(\sqrt{\tau}e^{y_q},\mu_1^2)c_{q_1(V_1)}^{pol}\right)
\cdot\left(\sum_{q_2(V_2)}f_{q_2}^{h_2}(\sqrt{\tau}e^{-y_q},
\mu_2^2)c_{q_2(V_2)}^{pol}\right)\cr
&&\quad\quad\cdot{\cal L}_{pol}\left(\hat{x},\sqrt{\frac{x}{\tau}}e^{y-y_q},
\frac{M_1^2}{s_{qq}},\frac{M_2^2}{s_{qq}}\right)
\quad + \quad h_1\leftrightarrow h_2\;\;\Bigg]\cr
&&\cdot\int\limits_{z_{\mathrm{min}}(y)}^{z_{\mathrm{max}}(y)}d\cos\theta
\frac{d\sigma}{d\cos\theta}((V_1 V_2)_{pol}\to V_3 V_4,{\cal W}^2),
\label{sigpp_cut}\end{eqnarray}
where the integration limits are determined by the rapidity-cut,
\begin{eqnarray}
y_{\mathrm{max}}&=&\min\left[Y,\frac{1}{2}\left(\frac{1}{x}\right)\right],\cr
z_{\mathrm{min}}(y)&=&\max\left[\frac{-\tanh(Y+y)}{\beta(M_3^2,M_4^2)},
\frac{-\tanh(Y-y)}{\beta(M_4^2,M_3^2)},-\cos\theta_{\mathrm{min}}\right],\cr
z_{\mathrm{max}}(y)&=&\min\left[\frac{\tanh(Y-y)}{\beta(M_3^2,M_4^2)},
\frac{\tanh(Y+y)}{\beta(M_4^2,M_3^2)},\cos\theta_{\mathrm{min}}\right],
\label{ycut_limits}\end{eqnarray}
with
\begin{eqnarray}
\beta(M^2,{M'}^2)&\equiv&\frac{\sqrt{1-\frac{2}{{\cal W}^2}(M^2+{M'}^2)
+\frac{1}{{\cal W}^4}(M^2-{M'}^2)^2}}{1+\frac{M^2-{M'}^2}{{\cal W}^2}}\cr
&=&\frac{q}{\sqrt{q^2+M^2}},
\end{eqnarray}
and  $\cos\theta_{\mathrm{min}}=1$. 
In the vicinity of the threshold for the production of the 
pair $V_3V_4$, the rapidity-cut has no effect anymore, i.e.
$z_{\mathrm{max}}(y)$ and $z_{\mathrm{min}}(y)$ 
are determined by $\cos\theta_{\mathrm{min}}=1$.

If the masses of the vector-bosons 
$V_3$ and $V_4$ are equal or only slightly different,
$M_3^2\simeq M_4^2$, or if the momenta of the bosons are large against
their masses, $q^2\gg\max(M_3^2,M_4^2)$, the expressions
(\ref{ycut_limits}) for $z_{\mathrm{min}}$ and $z_{\mathrm{max}}$
simplify to give
$z_{\mathrm{max}}=-z_{\mathrm{min}}=z_0$,
where
\begin{equation}
z_0=\min\left[\frac{\tanh(Y-|y|)}{\beta(M_3^2,M_4^2)},\cos\theta_{
\mathrm{min}}\right].
\label{z0hi}\end{equation}

If one applies a rapidity cut 
to the expression for convolutions of vector-boson distributions,
(\ref{lumiconv2}) with (\ref{vecinp2}),
one obtains the expression
\begin{eqnarray}
&&\frac{d\sigma}{dx}(h_1 h_2\to V_1 V_2\to V_3 V_4,s_{hh})|_{\mathrm{Cut}}\cr 
&=&\sum_{(V_1 V_2)}C_{(12)}\,\eta_0\sum_{pol=\lambda_1\lambda_2}
\int\limits_{-y_{max}}^{y_{max}}dy
\left[f_{V_{1,\lambda_1}}^{p_1}
(\sqrt{x}e^{y},\mu_1^2)f_{V_{2,\lambda_2}}^{p_2}
(\sqrt{x}e^{-y},\mu_2^2)
\;\;+\;\;h_1\leftrightarrow h_2\;\right]\cr
&&\quad\quad
\quad\quad\quad\quad\quad\quad\quad
\cdot\int\limits_{z_{\mathrm{min}}(y)}^{z_{\mathrm{max}}(y)}d\cos\theta
\frac{d\sigma}{d\cos\theta}((V_1 V_2)_{\lambda_1\lambda_2}
\to V_3 V_4,{\cal W}^2),
\label{sigtung_ycut}\end{eqnarray}
with $y_{max},z_{\mathrm{min}}(y)$ 
and $z_{\mathrm{max}}(y)$ from (\ref{ycut_limits}).

We calculate the differential cross-section
$d\sigma/dM_{ZZ}$ from Eq. (\ref{sigpp_cut}) with the
luminosities of the improved EVBA. As in \cite{ardv},
the quark-distributions
of EHLQ, \cite{EHLQ}, set 2, are used and the electroweak parameters are
$\alpha=1/128,s_W^2=0.22,M_W=80$ GeV, $M_H=0.5$ TeV and $\Gamma_H=51.5$ GeV.
For the scales $\mu_i^2$ in the quark-distributions,
$\mu_i^2=s_{qq}/4$ is chosen.
We also carry out a calculation with the convolutions
of vector-boson distributions from Eq. (\ref{sigtung_ycut})\footnote{
The value of $\mu_i^2$ in
$f_{q_i}^{h}(\xi_i,\mu_i^2)$ was again $\mu_i^2=\xi_is_{hh}$.}.

\begin{figure}
\begin{center}
\epsfig{file=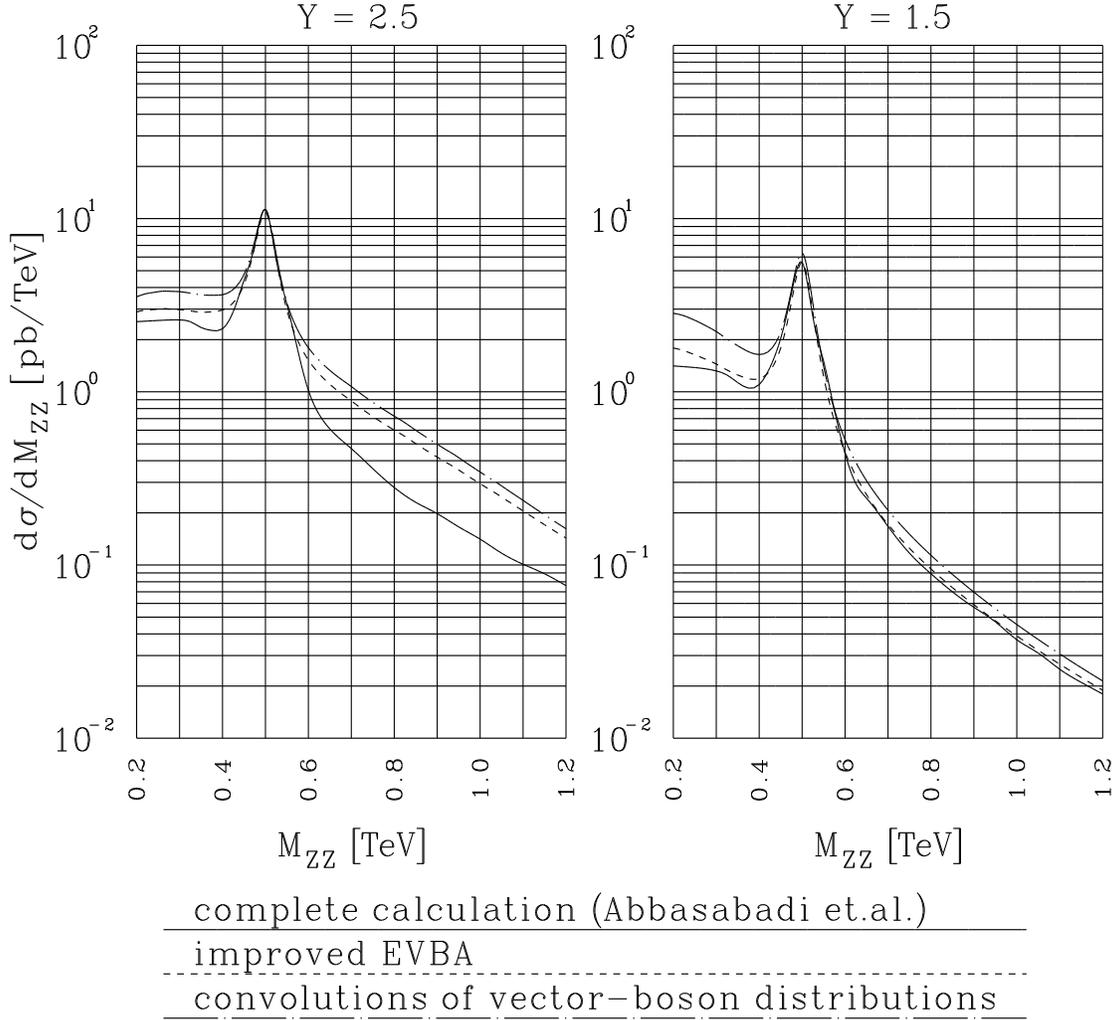,width=15cm,height=20cm}
\end{center}
\caption{The cross-section for $\protect pp\to ZZ X$ via
$W^+W^-$-scattering as a function of the invariant mass
$ M_{ZZ}$ at $\protect \sqrt{s_{hh}}=40$ TeV. A rapidity
cut of $Y=2.5$ and $Y=1.5$ was applied.
Shown is the result of the complete perturbative 
calculation \protect\cite{ardv}, the improved EVBA (\protect\ref{sigpp_cut})
and the result of the convolutions of vector-boson distributions
(\protect\ref{sigtung_ycut}).}
\label{comp}
\end{figure}

Figure \ref{comp} shows the cross-section for
$pp\to W^+W^-\to ZZ$ at a scattering energy of $\sqrt{s_{hh}}=40$ TeV
as a function of the invariant mass $M_{ZZ}$ of the $ZZ$ pair for
rapidity cuts of $Y=2.5$ and $Y=1.5$ as a result of the improved
EVBA calculation and the calculation with convolutions of
vector-boson distributions together
with the complete result from \cite{ardv}. 
For $Y=2.5$, the cross-section of the improved EVBA
deviates by a factor of two from the complete result
at $M_{ZZ}\gsim 0.7$ TeV. 
The result obtained with the convolutions deviates by $13\%$ ($M_{ZZ}=1.2$ TeV)
and $18\%$ ($M_{ZZ}=0.6$ TeV) from the improved EVBA result, independently
of the magnitude of the cut.  
For $Y=1.5$, a good agreement between the
improved EVBA and the complete calculation is found.
The EVBA deviates by less than $10\%$ from the exact result for
$M_{ZZ}>0.4$ TeV.

An explanation for the different results for $Y=2.5$ and $Y=1.5$
is that the brems\-strah\-lung-type diagrams in Fig. \ref{fig3}
begin to play a role if the angle between the produced vector-boson and
the hadron beam-direction is small.
This is the case for $Y=2.5$. In contrast,
the bremsstrahlung-diagrams might be neglected if only large angles 
are involved. This is the case for $Y=1.5$.
For a cut of $Y=2.5$ the smallest allowed angle is
$\theta_{\mathrm{min}}=9.4^{\circ}$, while the smallest angle
for $Y=1.5$ is $\theta_{\mathrm{min}}=25.2^{\circ}$.

In summary, we have seen that the improved EVBA deviates
by only ${\cal O}(10\%)$ 
from the result of a complete perturbative calculation
for a cut of $Y= 1.5$.
This result was found for $pp\to ZZ+X$ at
$\sqrt{s_{hh}}=40$ TeV and invariant masses of $\sqrt{{\cal W}^2}>0.4$ TeV.
There is, however, no reason
that a similar conclusion could not also be drawn for
the production of other vector-boson pairs, 
$pp\to V_3V_4+X$. We expect this because the
EVBA only pertains to the process-independent vector-boson luminosities.
The use of convolutions instead of the improved EVBA
leads to an additional error of $<20\%$ for $\sqrt{{\cal W}^2}\gsim 0.5$ TeV
($10\%$ at $\sqrt{{\cal W}^2}=2$ TeV).


\begin{figure}
\begin{center}
\epsfig{file=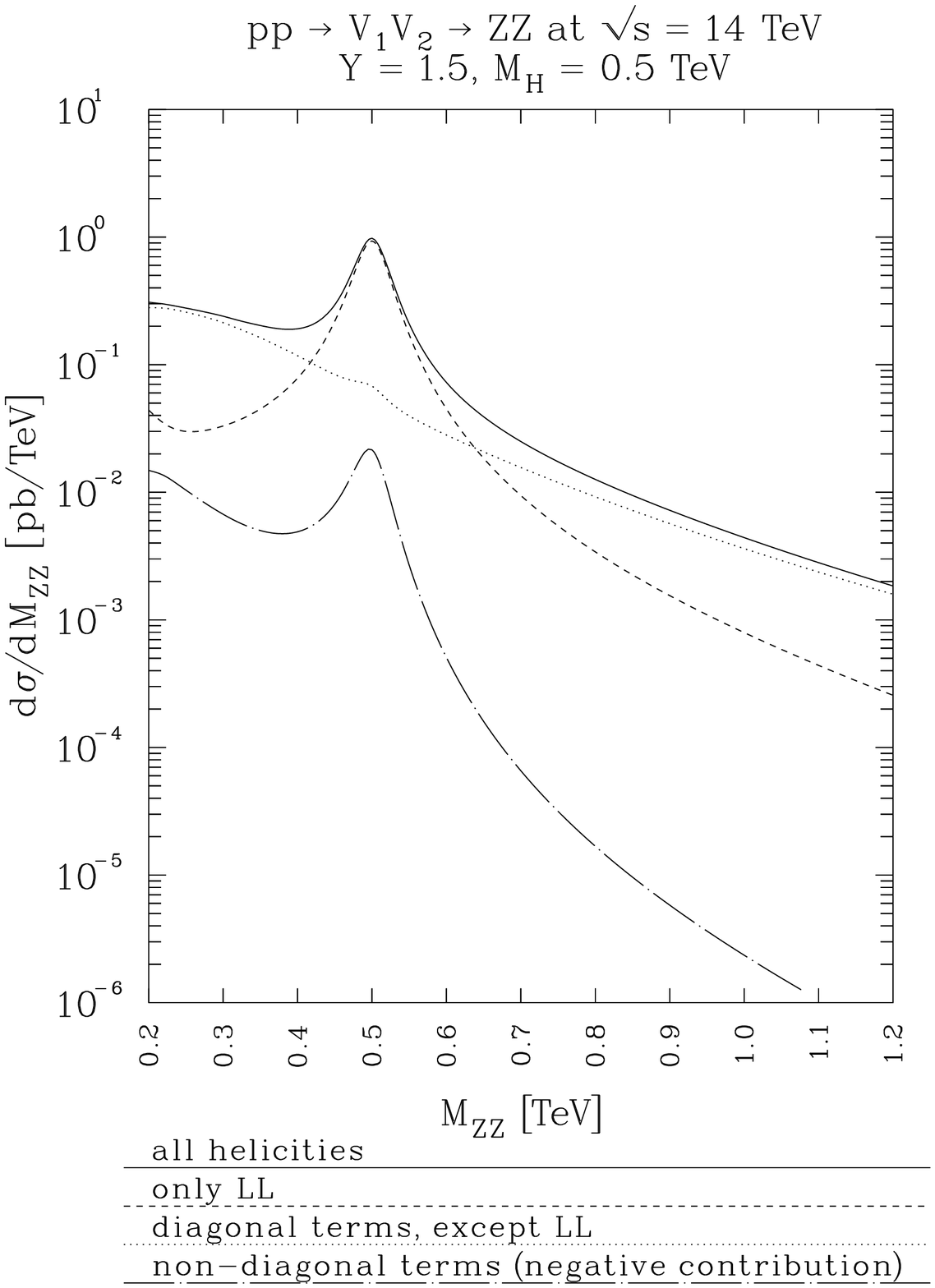,width=15cm,height=20cm}
\end{center}
\caption{The cross-section for $\protect pp\to (W^+W^- + ZZ)\to ZZ$ in the
improved EVBA, (\ref{sigpp_cut}), for a scattering energy of 
$\protect\sqrt{s_{hh}}=14$ TeV with
a rapidity cut of $Y=1.5$ as a function of the invariant mass
$M_{ZZ}$. The contribution from the $LL$-, the diagonal (without $LL$),
the non-diagonal and the sum of all
helicity combinations are shown separately.}
\label{ppzz}
\end{figure}

We finally present another result which is of interest in connection
with the EVBA. It concerns the magnitude of the off-diagonal terms
in the helicities of $V_1$ and $V_2$, denoted by $TTTT$, $TLTL$
and $\overline{T}L\overline{T}L$ in \cite{lumis}.
Figure \ref{ppzz} shows the contributions of the $LL$-, the other diagonal,
the non-diagonal and the sum of all
helicity combinations for the cross-section
for $pp\to (W^+W^- +ZZ)\to ZZ$ at $\sqrt{s_{hh}}=14$ TeV as a function
of the invariant mass $M_{ZZ}$ for a rapidity cut of $Y=1.5$.
The parameters and parton distributions were chosen as in Section
\ref{sec1} and the parameters for the Higgs boson were
$M_H=500$ GeV and $\Gamma_H=51.5$ GeV.
The sum of the non-diagonal helicity contributions, $TTTT$, $TLTL$
and $\overline{T}L\overline{T}L$, is negative and very small compared to the
diagonal helicity combinations. The non-diagonal terms can therefore
be safely neglected for this process. 
The longitudinal helicity combination, $LL$,
only plays a role near the Higgs resonance and is otherwise also
small. Important for the production of vector-boson pairs with large
invariant masses are the transverse helicities.


\section*{Conclusion}
We have given exact results for luminosities of vector-boson pairs
in a proton pair. 
In contrast to previous results, our treatment
of the effective vector-boson method made no
approximation in the integration over the phase space of the two
intermediate vector-bosons. 
The full calculation is involved but we have shown that
approximate expressions exist which reproduce the exact luminosities
to a fairly good degree. 
Identifying in detail the
approximations leading to the simple formalism of convolutions of
vector-boson distributions in hadrons we have
given a direct approximation to
the exact luminosities.
For one of the phenomenologically
interesting processes of vector-boson pair production in high-energy
proton-proton collisions we have shown that the direct approximation
deviates by less than $20\%$
from the result obtained with the exact luminosities.

In a numerical comparison of the improved EVBA with a complete
perturbative calculation for the process $pp\to ZZ+X$ we have shown
that the improved EVBA can reproduce the complete
result to ${\cal O}(10\%)$ if a rapidity cut of large enough strength is 
applied. This is true not only on the Higgs boson resonance but also
far away from it.
The improved EVBA thus gave a good approximation not only for longitudinal
but also for transverse vector-boson scattering. 
If a light Higgs boson exists,
this latter process is the dominating
production mechanism of high-energy vector-boson pairs in $pp$-collisions
at LHC energies.

We further investigated previous formulations of the EVBA.
These formulations always used the approximation of convolutions
of distribution functions of single vector-bosons.
We investigated  in detail the approximations which were made
and discussed the differences between various existing
derivations. Only some of
the derivations use no other approximations than
those inherent in the EVBA. We numerically addressed the deviation
of existing formulations from the exact luminosities.
The deviations are in general
larger than those of the direct approximation given here.

\section*{Acknowledgement}
I thank S. Dittmaier, D. Schildknecht and H. Spiesberger for
useful discussions.

\begin{appendix}
\section{Differences between Vector-Boson Distribution\\ 
Functions in the Literature\label{appa}}
In the main text I gave some results obtained by using vector-boson
distribution functions in fermions, $f_{V_{\lambda}}^{q}(\hat{z})$.
In this Appendix I specify the explicit forms which I used for the
functions and briefly discuss the differences of various
functions which were derived in the literature.

Vector-boson distribution functions have been derived by several
authors
\cite{dawson,karero,lindfors,capdequi,tung1,ardv,rolnick,godbole}.
All distributions describe the emission of a vector-boson 
as shown in Figure \ref{qqdia} according to Eq. (\ref{qqcross}).
In general the functions differ from each other because different
approximations and assumptions were made.
A discussion of the differences can be found in \cite{diss}.
We repeat the main points. 
In \cite{karero,ardv}, kinematic approximations concerning the
transverse momentum $k_{\perp}$ of the vector-boson were made.
It was assumed that $k_{\perp}^2\ll s_{qq}$. These approximations
were removed in \cite{rolnick}. Also in 
\cite{dawson,capdequi,godbole}, no approximations of this kind were
made. The distributions \cite{dawson,capdequi,godbole}
are all very similar to each other.
They have in common that the scale variable $\hat{z}$ was defined
as the ratio of the vector-boson's energy and the energy of the quark
$q$, $\hat{z}\equiv k^0/E$. For clarity, we define a single set of
distribution functions instead of using any particluar one
of the parametrizations 
\cite{dawson,capdequi,godbole} or \cite{rolnick}. The three parametrizations 
\cite{dawson,capdequi,godbole} all agree
if they are written in the form
\begin{equation}
f_{V_{\lambda}}^q(\hat{z})=\frac{\alpha}{2\pi}
\int\limits_{-4E^2(1-\hat{z})}^0 dk^2 \frac{T_{\lambda}}{(k^2-M^2)^2}
\frac{F_{Vp}(M^2)}{F_{lp}}\frac{|\hat{\cal M}|_\lambda^2(k^2)}
{|\hat{\cal M}|_\lambda^2(M^2)}.
\label{dgc1}\end{equation}
In (\ref{dgc1}),
$T_{\lambda}$ is the fermionic trace tensor contracted with
the polarization vectors $\epsilon(h)$,
\begin{eqnarray}
T_T&=&(v^2+a^2)\sum_{h=+,-}\left[l\cdot\epsilon^\ast(h)
l'\cdot\epsilon(h)+l\cdot\epsilon(h)l'\cdot\epsilon^\ast(h)+l\cdot l'
\right]\nonumber\\
T_{\overline{T}}&=&(2va)\sum_{h=+,-}(-1)^hi\epsilon^{\mu\mu'\rho\sigma}
l_\rho l'_\sigma\epsilon_\mu^\ast(h)\epsilon_{\mu'}(h)\nonumber\\
T_L&=&(v^2+a^2)\left[2l\cdot\epsilon(0)l'\cdot\epsilon(0)
-l\cdot l'\epsilon(0)\cdot\epsilon(0)\right].
\label{Tdef}\end{eqnarray}
The index $h$ is the helicity of the vector-boson and we used the
four-momenta defined in Figure \ref{qqdia}. $F_{Vp}(M^2)$ and
$F_{lp}$ are the on-shell flux factors for the scattering of the vector-boson
with the quark $q_2$ and for the scattering of the two quarks with each
other, respectively.
In terms of the particles' four-momenta the flux factors are given by
\begin{equation}
F_{Vp}(k^2)=\sqrt{(k\cdot p)^2+k^2p^2},\quad\quad
F_{lp}=\sqrt{(l\cdot p)^2+l^2p^2}.
\end{equation}
The quantities $|\hat{\cal M}|^2_{\lambda}(M^2)$ and
$|\hat{\cal M}|^2_{\lambda}(k^2)$ are the on-shell and off-shell, respectively,
squared
matrix elements for the scattering of the vector-boson with the quark $q_2$.

The polarization vectors were defined in a system 
in which the vector-boson
has a three-momentum $\vec{k}$ of magnitude $K$ along a particular direction
in space,
$\vec{k}=K\hat{e}_z$. We have $K^2=\hat{z}^2E^2-k^2$. 
Inserting the polarization vectors one obtains
\begin{eqnarray}
T_T&=&(v^2+a^2)\frac{(-k^2)\left[1+(1-\hat{z})^2-\frac{k^2}{2E^2}\right]}
{\hat{z}^2-\frac{k^2}{E^2}},\nonumber\\
T_{\overline{T}}&=&(2va)\frac{E}{K}(2-\hat{z})(-k^2).
\end{eqnarray}
To evaluate $T_L$, the polarization-vector $\epsilon(0)$ for an on-shell
vector-boson was used in \cite{dawson,godbole} while $\epsilon(0)$
for a vector-boson of arbitrary $k^2$ was used in \cite{capdequi}.
Since one has to integrate over $k^2$,
we use the latter choice, leading to
\begin{equation}
T_L=2(v^2+a^2)E^2(-k^2)\frac{1-\hat{z}+\frac{k^2}{4E^2}}{K^2}.
\end{equation}
So far, no reference
has been made to a specific frame.
In all distributions \cite{dawson,capdequi,godbole} the flux factor ratio
in (\ref{dgc1}) was
evaluated in the laboratory system of the quark $q_2$, thus,
\begin{equation}
F_{Vp}(M^2)/F_{lp}=\sqrt{\hat{z}^2-M^2/E^2}.
\end{equation}
In this frame, however, the 
relation $M_Y^2/s_{qq}=\hat{z}$ is only an approximate one.
It is really given by $M_Y^2/s_{qq}=\hat{z}+k^2/(2Em_q)$, 
where
$m_q$ is the mass of the quark $q_2$. I note that 
since the integration variable
$|k^2|$ becomes as large
as $4E^2(1-\hat{z})$, the desired connection between $\hat{z}$ and 
$M_Y^2/s_{qq}$ becomes completely disturbed even if $|k^2|$ is 
not even very large. 
It is therefore not meaningful to carry out the integration
over $k^2$ in the laboratory frame. The relation
$\hat{z}=M_Y^2/s_{qq}$, however, holds exactly in the cms 
of $q_1$ and $q_2$. We therefore evaluate the flux factor ratio in the
cms,
\begin{equation}
F_{Vp}(M^2)/F_{lp}=\hat{z}-\frac{M^2}{4E^2},
\end{equation}
and  
we have $4E^2=s_{qq}$. The remaining task in evaluating (\ref{dgc1})
is to make a model assumption about the $k^2$-dependence of the 
$|\hat{\cal M}|_\lambda^2$. The most simple assumption,
\begin{equation}
\frac{|\hat{\cal M}|_\lambda^2(k^2)}{|\hat{\cal M}|_\lambda^2(M^2)}= 1,
\label{assum1}\end{equation}
was made for all $\lambda$ in \cite{dawson,godbole}.
In \cite{capdequi}, more refined assumptions were made. 
These led to
the same simple relation (\ref{assum1}) 
for the $\lambda=T$ and different relations for
$\lambda=\overline{T}$ and $\lambda=L$. We will adopt here the follwoing 
minimal model assumptions,
\begin{eqnarray}
\frac{|\hat{\cal M}|_\lambda^2(k^2)}{|\hat{\cal M}|_\lambda^2(M^2)}&=&1,
\quad\lambda=T,\overline{T},\nonumber\\
\frac{|\hat{\cal M}|_\lambda^2(k^2)}{|\hat{\cal M}|_\lambda^2(M^2)}&=&
\frac{M^2}{-k^2},
\quad\lambda=L.
\label{offshell}
\end{eqnarray}
The same assumptions have been made in \cite{tung1}. They amount to
taking into account the $k^2$-dependence of the polarization vectors
and assuming no $k^2$ dependence otherwise.
The distribution functions are now given by
\begin{eqnarray}
f_{V_T}^q(\hat{z})=\frac{\alpha}{2\pi}(v^2+a^2)\left(\hat{z}-\frac{M^2}{4E^2}
\right)\int\limits_{-4E^2(1-\hat{z})}^0dk^2
\frac{(-k^2)\left[1+(1-\hat{z})^2-\frac{k^2}{2E^2}\right]}
{(k^2-M^2)^2\left(\hat{z}^2-\frac{k^2}{E^2}\right)},\nonumber\\
f_{V_{\overline T}}^q(\hat{z})=\frac{\alpha}{2\pi}(2va)
\left(\hat{z}-\frac{M^2}{4E^2}\right)
\int\limits_{-4E^2(1-\hat{z})}^0dk^2\frac{(-k^2)(2-\hat{z})}
{(k^2-M^2)^2\sqrt{\hat{z}^2-\frac{k^2}{E^2}}},\nonumber\\
f_{V_L}^q(\hat{z})=\frac{\alpha}{\pi}(v^2+a^2)
\left(\hat{z}-\frac{M^2}{4E^2}\right)M^2
\int\limits_{-4E^2(1-\hat{z})}^0dk^2\frac{1-\hat{z}+\frac{k^2}{4E^2}}
{(k^2-M^2)^2\left(\hat{z}^2-\frac{k^2}{E^2}\right)}.
\label{integrals}
\end{eqnarray}
The distribution function $f_{V_T}^q$ in (\ref{integrals}) is the
one given (in integrated form)
in \cite{dawson,godbole} (and it is also the one in \cite{capdequi},
but there are errors in the formulae given there) provided one 
divides these latter functions
by the flux factor ratio in the laboratory frame and multiplies
by the flux factor ratio in the cms, thus
\begin{equation}
f_{V_{\lambda}}^q=\frac{1-\frac{\textstyle M^2}
{\textstyle 4\hat{z}E^2}}{\sqrt{1-\frac{M^2}{\hat{z}^2E^2}}}
f_{V_{\lambda}}^q\left|{~}_{
\mbox{\footnotesize Literature\normalsize}}\right..
\label{fvqlit}\end{equation}
The distribution function $f_{V_{\overline T}}^q$ in (\ref{integrals})
is the one given in \cite{godbole} provided one applies
the same multiplication (\ref{fvqlit}).
The function $f_{V_{\overline T}}^q$
has only been given for $\hat{z}>M/E$ in \cite{godbole}.
For arbitrary values of $\hat{z}$ (in the allowed range 
$M^2/(4E^2)<\hat{z}<1$)
it is given by
\begin{equation}
f_{V_{\overline{T}}}^q(\hat{z})=\frac{\alpha}{2\pi}(2va)
\left(\hat{z}-\frac{M^2}{4E^2}\right)(2-\hat{z})I_3.
\label{fvtb}\end{equation}
The integral $I_3$ in (\ref{fvtb}) is defined by
\begin{equation}
I_3=\int\limits_{-4E^2(1-\hat{z})}^0\frac{dk^2(-k^2)}
{(k^2-M^2)^2\sqrt{\hat{z}^2-\frac{k^2}{E^2}}}.
\end{equation}
For $\hat{z}<M/E$, the result of the integration is
\begin{eqnarray}
I_3&=&-\frac{1}{\hat{z}^2-\frac{M^2}{E^2}}\left(
\hat{z}-\frac{M^2(2-\hat{z})}{4E^2(1-\hat{z})+M^2}\right)\nonumber\\
&&+\frac{2\hat{z}^2E^2-M^2}{2(\hat{z}^2E^2-M^2)}
\frac{2}{\sqrt{\frac{M^2}{E^2}-\hat{z}^2}}\left[
\arctan\left(\frac{2-\hat{z}}{\sqrt{\frac{M^2}{E^2}-\hat{z}^2}}\right)
-\arctan\left(\frac{\hat{z}}{\sqrt{\frac{M^2}{E^2}-\hat{z}^2}}\right)
\right].
\end{eqnarray}
This result thus continues the result given in \cite{godbole}
into the region $\hat{z}<M/E$.
The distribution function $f_{V_L}^q$ in (\ref{integrals}) is 
different from any one of those in \cite{dawson,capdequi,godbole}.
It is given by
\begin{equation}
f_{V_L}^q(\hat{z})=\frac{\alpha}{\pi}(v^2+a^2)
\left(\hat{z}-\frac{M^2}{4E^2}\right)
\left[(1-\hat{z})I_4-\frac{M^2}{4E^2}I_1\right],
\label{fvl}\end{equation}
with the integrals
\begin{eqnarray}
I_1&=&E^2\int\limits_{-4E^2(1-\hat{z})}^0\frac{dk^2(-k^2)}
{(k^2-M^2)^2(\hat{z}^2E^2-k^2)}\nonumber\\
&=&\frac{\hat{z}^2E^4}{(\hat{z}^2E^2-M^2)^2}
\left[\ln\left(\frac{4E^2(1-\hat{z})+M^2}{M^2}\right)
-2\ln\frac{2-\hat{z}}{z}\right]\nonumber\\
&&-\frac{E^2}{\hat{z}^2E^2-M^2}\left\{1-\frac{M^2}{4E^2(1-\hat{z})+M^2}
\right\},\\
I_4&=&M^2E^2\int\limits_{-4E^2(1-\hat{z})}^0\frac{dk^2}
{(k^2-M^2)^2(\hat{z}^2E^2-k^2)}\nonumber\\
&=&\frac{E^2}{\hat{z}^2E^2-M^2}\left(1-\frac{M^2}{4E^2(1-\hat{z})+M^2}
\right)\nonumber\\
&&-\frac{M^2E^2}{(\hat{z}^2E^2-M^2)^2}\left[
\ln\left(\frac{4E^2(1-\hat{z})+M^2}{M^2}\right)
-2\ln\frac{2-\hat{z}}{\hat{z}}\right]
\label{int}.
\end{eqnarray}
The distribution functions (\ref{integrals}) are defined for all values
of $\hat{z}$ in the range 
\begin{equation}
M^2/s_{qq}<\hat{z}<1,
\label{zhatlimits}\end{equation}
where $s_{qq}=4E^2$, and they are zero otherwise. 
The lower limit in (\ref{zhatlimits}) is meaningful because
the cross-sections for on-shell vector-boson scattering vanish
for $\hat{z}s_{qq}=M_Y^2<M^2$.
We will use the distributions
(\ref{integrals}) in the
main text and sometimes refer to them as DGC.

Having evaluated the functions (\ref{dgc1}) in the center-of-mass frame
of the quarks
we have induced an additional approximation, namely that the helicities
$h=0,\pm 1$ are not well-defined. To the order $k_{\perp}^2/E^2$, there
appears mixing between the helicity states. In particular, the transverse
and longitudinal helicity states mix. 
To see this we note that the on-shell cross-section 
$\sigma(V_{1,\lambda_1}q_2\to Y,M_Y^2)$ appearing in (\ref{qqcross})
must be evaluated for definite values of the components of the
four-vectors $k$ and $p$ since one has to use specific polarization
vectors. In particular, the components can not
depend on the integration variable $k^2$ appearing in (\ref{integrals}).
Of course, for a given value of the integration variable, a 
Lorentz-transformation into a frame in which $k$ and $p$ have given
components may be applied. However, this transformation
in general changes the helicity of the vector-boson.
Only in frames which are related to each other by a
boost in the direction of motion of the vector-boson the helicity
is the same. Therefore, in the frame in which the helicity is
defined, the transverse components of $p$ with
respect to $k$ must be the same for all values of the integration
variable $k^2$. For the distributions (\ref{integrals}) 
evaluated in the $q_1q_2$ center-of-mass frame this
is not the case. I note that the helicity could have been defined
without an approximation in the laboratory frame. It thus seems that with the
distributions (\ref{integrals}) we can choose between either having
mixing of the helicity states or a violation of the relation
(\ref{zhatdef}).

The above-mentioned approximations were avoided in the derivations \cite{lindfors,tung1}.
By defining $\hat{z}$ directly as $\hat{z}\equiv M_Y^2/s_{qq}$
(i.e. not as a ratio of energies) and defining the vector-boson helicity in its
Breit frame no approximations of kinematic origin were made.
The only remaining necessary (in the framework of the EVBA)
assumption concerned
the continuation of the vector-boson cross-sections into the
region of virtual vector-bosons. 
In \cite{lindfors}, the specific assumption that the final state $Y$
couples like a fermion to the intermediate vector-boson was made.
In \cite{tung1}, the minimal assumptions (\ref{offshell}) were used.
Concerning \cite{lindfors},
I note that the expression for the integral $I_2(\hat{z})$ 
given there is not correct. 
This expression would lead to vector-boson distribution
functions which become infinite as $\hat{z}\to M^2/s_{qq}$.
The expression must be replaced by
\begin{equation}
I_2(\hat{z})=[a^2+2\hat{z}r(1-r)a]\ln\left(\frac{a}{\hat{z}r}\right)
+\ln(\hat{z})+b(1-2a),
\end{equation}
where I used the variables $r,a$ and $b$ defined in \cite{lindfors}.
Concerning \cite{tung1}, I note that the flux factor $F_{Vp}$ was evaluated at
$k^2=0$ in \cite{tung1} but it should 
be evaluated at $k^2=M^2$
(since it is the cross-section for on-shell vector-bosons which appears in
(\ref{qqcross})). I therefore multiplied the distributions of
\cite{tung1} by the flux factor ratio 
$F_{Vp}(M^2)/F_{Vp}(0)=(1-M^2/(\hat{z}s_{qq}))$
before using them for numerical examples. 
Like the distributions (\ref{integrals}),
the distributions \cite{lindfors,tung1} are defined for all values of 
$\hat{z}$ in the range (\ref{zhatlimits}) and they are zero otherwise.

All distribution functions reduce to the same analytical expressions
if a crude approximation is made. 
This approximation is obtained
by retaining only the leading terms in the limit of vanishing vector-boson
masses, $M^2\ll \hat{z}s_{qq}$ and $M^2\ll (1-\hat{z})\hat{z}s_{qq}$.
This approximation has been
frequently used in the literature and has been called the leading
logarithmic approximation (LLA)\footnote{It should be noted that not all
leading terms are of logarithmic type.}.  
Expressions for the $f_{V_\lambda}^q$ in the LLA can be 
found e.g. in \cite{dawson,capdequi,lindfors2}. We use the lower
limit for $\hat{z}$,
$\hat{z}>M^2/s_{qq}$, also for the LLA distributions.
\end{appendix}

\end{document}